  \documentclass[structabstract]{aa}
\usepackage{graphicx}
\usepackage{txfonts}
\usepackage{natbib}
\usepackage{gensymb}
\usepackage{amsmath}

\bibpunct[]{(}{)}{;}{a}{}{,}

\begin{document}

   \title{The rotation-tunneling spectrum of 3-hydroxypropenal and confirmation of its 
          detection toward IRAS 16293$-$2422~B\thanks{Supplementary data are available at 
          CDS via anonymous ftp to cdsarc.u-strasbg.fr (130.79.128.5) or via 
          http://cdsweb.u-strasbg.fr/cgi-bin/qcat?J/A+A/}}

   \author{H.~S.~P. M{\"u}ller\inst{1}
           \and
           A. Coutens\inst{2}
           \and
           J.~K. J{\o}rgensen\inst{3}
           \and
           L. Margul{\`e}s\inst{4}
           \and
           R.~A. Motiyenko\inst{4}
           \and
           J.-C. Guillemin\inst{5}
           }

   \institute{Astrophysik/I.~Physikalisches Institut, Universit{\"a}t zu K{\"o}ln,
              Z{\"u}lpicher Str. 77, 50937 K{\"o}ln, Germany\\
              \email{hspm@ph1.uni-koeln.de}
              \and
              Institut de Recherche en Astrophysique et Plan{\'e}tologie (IRAP), 
              Universit{\'e} de Toulouse, UT3-PS, CNRS, CNES, 
              9 av. du Colonel Roche, 31028 Toulouse Cedex 4, France
              \and
              Niels Bohr Institute, University of Copenhagen, {\O}ster Voldgade 5$-$7, 1350 Copenhagen K, Denmark 
              \and
              Univ. Lille, PhLAM $-$ Physique des Lasers, Atomes et Mol{\'e}cules, 
              CNRS, UMR 8523, F-59000 Lille, France
              \and
              Univ Rennes, Ecole Nationale Sup{\'e}rieure de Chimie de Rennes, 
              CNRS, ISCR$-$UMR 6226, F-35000 Rennes, France
              }

   \date{Received 19 Mar 2024 / Accepted 30 Apr 2024}
 
  \abstract
{3-Hydroxypropenal (HOCHCHCHO) is the lower energy tautomer of malonaldehyde which displays 
a complex rotation-tunneling spectrum. It was detected somewhat tentatively toward the solar-type 
protostellar system IRAS 16293$-$2422 with ALMA in the framework of the Protostellar 
Interferometric Line Survey (PILS). Several transitions, however, had large residuals, 
preventing not only their detection, but also the excitation temperature of the species 
from being determined unambiguously.}
{We want to extend the existing rotational line list of 3-hydroxypropenal to shed more 
light on the recent observational results and to facilitate additional radio astronomical 
searches for this molecule.}
{We recorded and analyzed the rotation-tunneling spectrum of 3-hydroxypropenal in the frequency 
regions between 150 and 330~GHz and between 400 and 660~GHz. Transitions were searched for 
in the PILS observations of IRAS~16293$-$2422. Local thermodynamic equilibrium (LTE) models 
were carried out and compared to the observations to constrain the excitation temperature. 
Additional transitions were searched for in other ALMA archival data of the same source 
to confirm the presence of 3-hydroxypropenal.}
{More than 7500 different spectral lines, corresponding to more than 11500 transitions, 
were assigned in the course of our investigation with quantum numbers $2 \le J \le 100$, 
$K_a \le 59$, and $K_c \le 97$, resulting in a greatly improved set of spectroscopic 
parameters. The comparison between the LTE models and the observations yields an 
excitation temperature of 125~K with a column density $N = 1.0 \times 10^{15}$~cm$^{-2}$ 
for this species. We identified seven additional lines of 3-hydroxypropenal that show 
a good agreement with the model in the ALMA archive data.}
{The calculated rotation-tunneling spectrum of 3-hydroxypropenal has sufficient accuracy for 
radio astronomical searches. With the solution of the excitation temperature conundrum and 
the detection of seven more lines, we consider the detection of 3-hydroxypropenal toward 
IRAS 16293$-$2422 as secure.} 

\authorrunning{H.~S.~P. M{\"u}ller et al.}
\titlerunning{The rotation-tunneling spectrum of 3-hydroxypropenal}

\maketitle
\hyphenation{For-schungs-ge-mein-schaft}

\section{Introduction}
\label{intro}

\begin{figure*}
\centering
\includegraphics[height=3cm,angle=0]{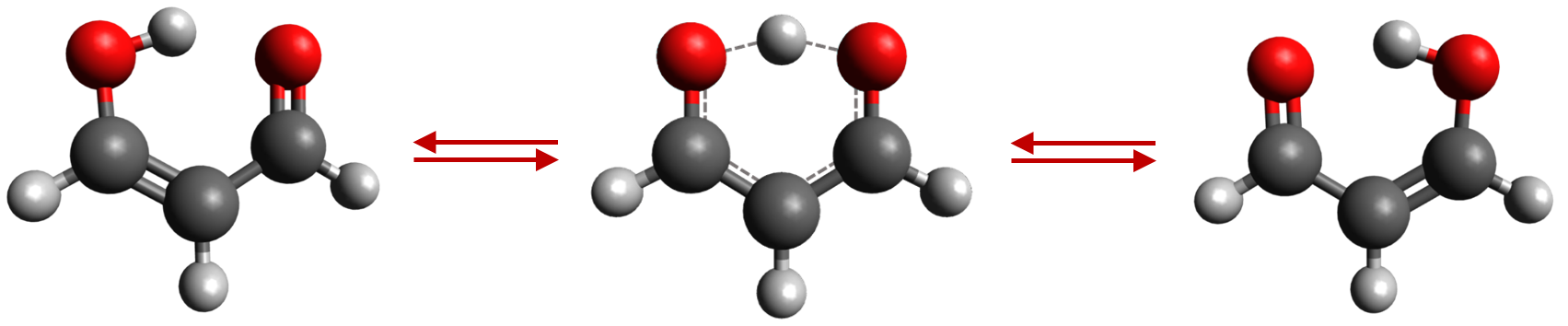}

\caption{Model of the 3-hydroxypropenal molecule. Carbon atoms are symbolized by gray spheres, 
   hydrogen atoms are indicated by small, light gray spheres, and oxygen atoms by red spheres. 
   The two equivalent minimum structures are shown at the outside while the transition state 
   with $C_{\rm 2v}$ symmetry is displayed in the center.}
\label{3-HP-structure}
\end{figure*}


Malonaldehyde is the prototype and parent species of $\beta$\nobreakdash-dicarbonyl compounds 
of the form RC(O)CH$_2$C(O)R' (R = R' = H for malonaldehyde). However, it was found that enol 
tautomers of these molecules were usually lower in energy because of the formation of 
intramolecular hydrogen bonds. An interesting question in the special case of R = R' was 
if this hydrogen bonding was strong enough to form symmetric hydrogen bonds to both oxygen atoms 
or if the bonding was weaker to form a hydrogen bond with only one oxygen atom, thus yielding a 
double minimum potential possibly with tunneling between the two minima. 
The enol tautomer of malonaldehyde is 3\nobreakdash-hydroxypropenal, one of the simplest 
molecules containing a ring which is closed by a hydrogen bond. Early quantum-chemical 
calculations yielded contradicting results on its structure; \citet{3-HP_ai_1975a} 
found a structure with an asymmetric double minimum potential with a barrier low enough 
to permit tunneling between the two minima, as presented in the left and right parts of 
Fig.~\ref{3-HP-structure}, whereas \citet{3-HP_ai_1975b} found a structure with a single, 
symmetric minimum potential, such as the one shown in the center of Fig.~\ref{3-HP-structure}.


The rotational spectra of several 3-hydroxypropenal isotopologs were investigated extensively 
in order to resolve this conflict. Two sets of transitions with similar spectroscopic parameters, 
but opposing spin-statistics (3\,:\,1 versus 1\,:\,3) were identified in an early, preliminary 
account \citep{3-HP_rot_early-account_1976}. The second set belonged to a state $16 \pm 14$~cm$^{-1}$ 
higher than that of the first set, compatible with a double minimum potential with tunneling and 
almost certainly ruling out a symmetric single minimum potential. Comparison of the OO non-bonding 
distance, determined through $^{16}$O/$^{18}$O substitution, with values from quantum-chemical 
calculations \citep{3-HP_ai_1975a,3-HP_ai_1975b} supported this interpretation. 
\citet{3-HP_rot_etc_1981} investigated many more isotopic species, extended the measurements into 
the lower millimeter region, carried out dipole moment measurements, refined the energy difference 
of the two states to $26 \pm 10$~cm$^{-1}$ and observed an absorption band near 21~cm$^{-1}$ for 
the main isotopolog, and determined the structural parameters of the asymmetric minimum configuration 
of 3-hydroxypropenal. Two later studies \citep{3-HP_analysis_1_1984,3-HP_analysis_2_1984} improved 
the analyses of the rotation-tunneling interaction and reported additional transition frequencies 
for the main isotopolog \citep{3-HP_analysis_2_1984}. \citet{3-HP_analysis_1_1984} also evaluated 
the tunneling barrier height for the main isotopolog as $\sim$6.6~kcal/mol or $\sim$2300~cm$^{-1}$.

\citet{3-HP_rot_Zeeman_1983} studied the rotational Zeeman effect of 3-hydroxypropenal and 
reported field-free transition frequencies for some low-$J$ transitions. 
\citet{3-HP_FIR_1991} applied tunable far-infrared spectroscopy to extend transition frequencies of 
the main isotopolog into the submillimeter region. 
Finally, \citet{3-HP_tunneling_1999} reported several transition frequencies between the two 
tunneling states and evaluated the associated $a$-dipole moment component through relative intensity 
measurements in comparison to the rotational transitions within the tunneling states, which obey 
$b$-type selection rules.

Recently, \citet{3-HP_det_2022} employed a catalog entry of the Cologne Database for Molecular 
Spectroscopy, \citep[CDMS,][]{CDMS_2005,CDMS_2016} to identify 3\nobreakdash-hydroxypropenal 
somewhat tentatively toward the B component of the protostellar system IRAS 16293$-$2422 
(hereafter IRAS16293) in observations with the Atacama Large Millimeter/submillimeter Array (ALMA) 
obtained in the Protostellar Interferometric Line Survey (PILS), an unbiased molecular line survey 
between 329.1 and 362.9~GHz \citep{PILS_overview_2016}. 
The 3-hydroxypropenal catalog entry had one decisive drawback that transition frequencies 
in the frequency range of the survey with low values of $K_a$ had rather large uncertainties, 
preventing to establish their presence or absence. Such transitions should be observable in the 
warm excitation temperature scenario of $\sim$300~K, whereas they should be too weak to be 
identified in the luke-warm excitation scenario of $\sim$125~K. Both excitation scenarios were 
found quite commonly for molecules identified in PILS \citep[e.g.,][]{PILS_div-isos_2018}.

We have recorded and analyzed the rotation-tunneling spectrum of 3-hydroxypropenal in the millimeter 
and submillimeter region in order to establish its excitation temperature in IRAS16293~B, 
which may lead to greater certainty about the presence of the molecule in this source, and to 
facilitate further searches for 3-hydroxypropenal in the interstellar medium.

We provide in Section~\ref{rot_backgr} the spectroscopic properties of the molecule, in 
Section~\ref{exptl} the experimental details, describe in Section~\ref{lab-results} 
the laboratory spectroscopic results, and discuss these in Section~\ref{lab-discussion}. 
The astronomical observations are detailed in Section~\ref{ALMA-observations}, and 
conclusions and outlook are presented in Section~\ref{conclusion}.

\section{Spectroscopic properties of 3-hydroxypropenal}
\label{rot_backgr}

The 3-hydroxypropenal molecule is a very asymmetric rotor of the prolate type with 
$\kappa = (2B - A - C)/(A - C) = -0.4441$. It has two equivalent configurations of 
$C_{\rm S}$ symmetry, as displayed schematically in Fig.~\ref{3-HP-structure}.  
The barrier to interchange between these two minima is sufficiently low to facilitate tunneling, 
which leads to a symmetric tunneling state $\varv = 0^+$ and an antisymmetric tunneling state 
$\varv = 0^-$ 647046.2~MHz higher in energy \citep{3-HP_tunneling_1999}.  
The transition state of $C_{\rm 2v}$ symmetry is shown in the center of Fig.~\ref{3-HP-structure}. 
The two equivalent H atoms not on the symmetry axis in the transition state lead to 
\textit{ortho}/\textit{para} spin-statistics with relative intensities of 3\,:\,1. 
The \textit{ortho} levels in $\varv = 0^+$ are those with $K_a + K_c$ being odd and 
in $\varv = 0^-$ those with $K_a + K_c$ being even.

The dipole moment component for rotational transitions, within $\varv = 0^+$ or $\varv = 0^-$, 
is along the $b$-axis, the symmetry axis of the transition state. Its magnitudes were determined 
through Stark spectroscopy as $\mu _b = (2.59 \pm 0.02)$~D for $\varv = 0^+$ and 
$\mu _b = (2.58 \pm 0.02)$~D for $\varv = 0^-$ \citep{3-HP_rot_etc_1981}. The selection rules are 
$\Delta J = 0$, $\pm$1 and $\Delta K_a$ and $\Delta K_c$ are odd. The strongest transitions are 
$^rR$-branch transitions with $\Delta K_a = +1$ and $\Delta J = +1$ from the lower to the upper 
state and $K_a$ close to $J$. Such transitions were reported by \citet{3-HP_FIR_1991} in their 
tunable far-infrared study; transitions with very low values of $K_a$ were not reported. 
Also at lower $K_a$, $^pR$-branch transitions ($\Delta K_a = -1$) are relatively strong 
as they form asymmetry doublets with the $^rR$-branch transitions having the same $J$ 
and $K_c$. The respective pair of transitions is well separated at intermediate $K_a$ 
and $K_c$, but the splitting decreases rapidly with decreasing $K_a$ and is collapsed 
well before $K_a = 0$. This pairing of asymmetry doublets is called oblate pairing while 
the respective pairing at high values of $K_a$ is called prolate pairing. 
The $Q$-branch transitions ($\Delta J = 0$) are usually weaker, but higher-$J$ may appear as 
quite strong compared to lower-$J$ $R$-branch transitions, which is quite commonly found 
at lower frequencies. The $P$-branch transitions ($\Delta J = -1$) are usually weaker still. 
Finally, the asymmetry of the molecule causes transitions with $\Delta K_a = \pm3$ etc. 
to have non-negligible intensities.

The tunneling of the H atom is along the $a$-axis; transitions between $\varv = 0^+$ and 
$\varv = 0^-$ follow therefore $a$-type selection rules with $\Delta J = 0$, $\pm$1, 
$\Delta K_a$ even, and $\Delta K_c$ odd. Strong rotation-tunneling transitions are 
$^qQ$-branch transitions with $K_a$ close to $J$. 
The transitions reported by \citet{3-HP_tunneling_1999} were mostly of this type as they occur 
in a rather narrow frequency window quite close to the tunneling frequency. A small number of 
transitions with somewhat lower values of $K_a$ were not included in their fit because they were 
weak or blended. And again were transitions with very low values of $K_a$ not reported in that work. 
\citet{3-HP_tunneling_1999} determined $\mu_a/\mu_b = 0.14$ from relative intensity measurements. 
This yields $\mu_a = 0.36$~D; it was assumed that $\mu_a$ is positive, as is commonly done. 
The intensities of transitions involving at least one perturbed level may, however, be affected 
by the sign of $\mu_a$ relative to that of $\mu_b$ and that of $F_{ab}$, as will be discussed 
in Section~\ref{assignments}. 
Also relatively strong are $^qP$- and $^qR$-branch transitions having low values of $K_a$. 
The $^qR$-branch transitions have frequencies above the tunneling frequency for the most part 
while $^qP$-branch transitions are usually found below it. 
Transitions with $\Delta K_a = \pm2$ etc. gain considerable intensity because of the asymmetry of 
the 3-hydroxypropenal molecule. In addition, they may borrow additional intensity from rotational 
transition through tunneling-rotation interaction.

Energy levels in $\varv = 0^+$ and $\varv = 0^-$ with similar energies can interact if they 
have the same $J$ and if they differ in $K_a$ by an odd number and in $K_c$ by an even number. 
The interaction is a Coriolis-type interaction, frequently just called Coriolis interaction. 
It is usually strongest the closer the energies are in the unperturbed case, 
the smaller $\Delta K_a$ and $\Delta K_c$ are, and the higher $J$ and $K_a$ or $K_c$ are. 
Consequences of the interactions are that the two levels repel each other and mix in character 
with strength of the interaction. Transitions may be displaced by several gigahertz, and 
tunneling-rotation transitions can gain also intensity from rotational transitions through 
the mixing.

\section{Experimental details}
\label{exptl}

\subsection{Sample preparation}
\label{synthesis}

The malonaldehyde synthesis of \citet{MA-synth_2008} was applied with slight modifications. 
Acid hydrolysis of 1,1,3,3-tetraethoxypropane (purity > 96\%, Sigma Aldrich) followed 
by treatment with aqueous NaOH gave the sodium salt of malonaldehyde. 
This salt suspended in diethyl ether was then acidified with an anhydrous solution of HCl 
in diethyl ether at $-$40$^{\rm o}$C followed by 2~h of stirring at this temperature. 
Purification was carried out by trap-to-trap distillation with slow heating to room temperature 
of the solution to give pure malonaldehyde in a trap immersed in a bath cooled to $-$50$^{\rm o}$C.

\subsection{Spectroscopic measurements}
\label{lab-spec}

The measurements between 150 and 330~GHz and between 400 and 660~GHz were carried out with 
the Lille spectrometer \citep{Lille-spectrometer}. 
We employed a quartz tube (10~cm diameter, 200~cm in length) as absorption cell. 
Throughout the measurement, the sample was submerged in an ethanol cold bath at $-$15$^{\rm o}$C, 
and a minimum flow of the sample vapor was maintained between 2.0 to 2.5~Pa (20$-$25~$\mu$bar).
The frequencies were covered with various active and passive frequency 
multipliers from VDI Inc., and an Agilent synthesizer (12.5$-$18.25~GHz) was used as 
the source of radiation. A liquid He-cooled InSb bolometer (QMC Instruments Ltd) was used 
to detect the absorption signals. The uncertainties of the measured line position were judged 
on the symmetry of the line-shape and the signal-to-noise ratio (S/N); 15, 20, or 30~kHz were 
assigned to average lines, 10 and 5~kHz for very symmetric lines with very good S/N, and 
50, 70, or 100~kHz for weaker lines, lines with poorer S/N, or lines with less symmetric 
line-shape, for example, because of the proximity to another line.

\section{Laboratory spectroscopic results}
\label{lab-results}
\subsection{Analysis and fitting}
\label{analysis}

The rotation-tunneling spectrum, such as in the case of 3\nobreakdash-hydroxypropenal, 
is usually treated with a Hamiltonian that is divided into a $2 \times 2$ matrix. 
The diagonal elements are commonly two Watson-type rotational Hamiltonians and include 
the energy of the upper tunneling state; the interaction Hamiltonian is off-diagonal. 
We applied here Watson's S reduced Hamiltonian in the prolate I$^r$ representation on 
the diagonal; see \citet{18O-SO2_rot_AuS-red_2020} for a discussion on the advantages 
of the S reduction.

The two tunneling states $0^+$ and $0^-$ together comprise the ground vibrational state $\varv = 0$. 
Therefore, it can be advantageous to rearrange the Hamiltonians and fit average spectroscopic parameters 
$X = (X(0^+) + X(0^-$))/2 and differences $\Delta X = (X(0^-) - X(0^+$))/2, as was done by 
\citet{aGg-eglyc_rot_2003} in their treatment of the lowest energy conformer of ethylene glycol. 
\citet{aGg-eglyc_rot_2003} also pointed out that the differences in spectroscopic parameters can be 
interpreted as rotational corrections to the energy difference. We follow this interpretation in the 
present work. We should point out that in our definition of the differences, $0^-$ is higher in energy 
by $2E$ than $0^+$. The advantage of this formulation is that an average parameter or its difference 
can be used individually in the fit independent of each other and was particularly noteworthy in a 
refit of ethanethiol data \citep{RSH_ROH_2016}. Other investigations applying this approach include 
hydroxyacetonitrile \citep{HAN_rot_2017} and some of its minor isotopic species \citep{HAN_isos_rot_2023} 
as well as dimethylamine \citep{DMA_rot_2023}.

The off-diagonal interaction was described employing Pickett's Reduced Axis System (RAS) 
\citep{RAS_Pickett_1972}, as was already done in earlier studies of the rotation-tunneling 
spectrum of 3-hydroxypropenal. The only non-zero low-order term is $F_{ab}(J_aJ_b + J_bJ_a)$ 
that may be supplemented with suitable distortion parameters of the form 
($F_{ab,K}J_a^2 + F_{ab,J}J^2 + F_{2ab}(J_b^2 - J_c^2) + ...) \times (J_aJ_b + J_bJ_a)$.

Pickett's programs SPCAT and SPFIT \citep{spfit_1991} were used to calculate and fit the 
rotational spectra of 3\nobreakdash-hydroxypropenal. 
We determined the spectroscopic parameters of 3-hydroxypropenal in the usual way. 
We tested after each round of assignments if one or more spectroscopic parameters would 
improve the quality of the fit by amounts that warranted keeping the respective parameter 
in the fit. This procedure helps to keep the number of spectroscopic parameters small 
while reproducing the transition frequencies already in the fit as well as possible. 
It is important to try only parameters that are reasonable with respect to those 
already used in the fit. If at least one parameter improved the quality of the fit sufficiently, 
we chose the one that improved the quality the most and searched for additional parameters. 
We tested occasionally if a parameter with relatively large uncertainties can be omitted from 
the fit without increasing the rms error of the fit by too large amounts.

\subsection{Previous data}
\label{prev-data}

We describe the previous data, their treatment in the initial fit that was the basis of 
the version 1 CDMS catalog entry of 3\nobreakdash-hydroxypropenal, which in turn was the 
starting point of our present analysis, and their treatment in the final fit of the present study. 
The microwave data from \citet{3-HP_rot_etc_1981} up to 40~GHz displayed very small residuals 
and were assigned uncertainties of 10~kHz in the initial fit as well as in the final fit 
of this study since no uncertainties were given in that work. The $b$-type $^rR$- and 
$^rQ$-branch transitions cover quantum numbers $1 \le J \le 15$ and $K_a \le 6$. 
The millimeter wave transitions between 55 and 113~GHz were $^rQ$-branch transitions with 
$6 \le J \le 29$ and $2 \le K_a \le 11$ that were taken with a different spectrometer. 
Uncertainties of 150~kHz were applied in the initial fit and one transition with $J = 37$ 
was omitted because of large residuals. These data were omitted in the final fit of 
the present work because of transition frequencies with similar quantum numbers with much 
smaller uncertainties.

\citet{3-HP_rot_Zeeman_1983} published microwave transitions between 12 and 25~GHz with 
$K_a \le J \le 3$. They reported uncertainties as being better than 20~kHz. We included them 
in the initial fit as well as in the final fit from our current investigation with 10~kHz 
uncertainties.

\citet{3-HP_analysis_2_1984} reported a moderate number of $^rR$- and $^rQ$-branch transitions 
between 78 and 85~GHz having $8 \le J \le 38$ and $K_a \le 14$ with estimated uncertainties of 
200~kHz. The transition frequencies were used in the initial fit except for three. Their data 
were omitted from our present final fit for the same reason as above.

The tunable far-infrared data published by \citet{3-HP_FIR_1991} are $^rR$-branch transitions, 
except for two $^pR$-branch transitions, distributed between 352 and 898~GHz with $20 \le J \le 50$ 
and $10 \le K_a \le 44$. Uncertainties were reported to be about 400~kHz, which appeared to be 
too optimistic. We assigned 800~kHz in out initial fit and omitted three transition frequencies 
because of large residuals nevertheless. We omitted these data in our final fit because of the 
large uncertainties and because we have remeasured many of these transitions more accurately.

The only tunneling-rotation transitions were reported by \citet{3-HP_tunneling_1999}. 
They published $^qQ$-branch transitions between 643 and 651~GHz with $8 \le J \le 35$ 
and $4 \le K_a \le 30$. No uncertainties were given explicitely, but the rms of $\sim$30~kHz 
for their lines included in the fit was deemed to be commensurate with the uncertainties. 
We included the data in our initial fit accordingly, but omitted them in our final fit 
of this study. Not only did we redetermine frequencies of most transitions, often with better 
uncertainties, but we also noted that the series of transitions with $K_a = J - 4$ and 
$K_a = J - 5$ differed in many cases from our measurements and from the calculated frequencies 
by 50~kHz and more; only the short series of transitions with $K_a = J$ showed very good 
agreement with the exception of one line that was not included in their fit and was not 
assigned in the present work.

\subsection{Observed spectrum and assignments}
\label{assignments}

The recorded rotation-tunneling spectrum of 3-hydroxypropenal displayed fluctuations in the S/N, 
which is commonly observed in millimeter and submillimeter spectroscopy, but the change in S/N 
is mostly gradual. In combination with a rather rich spectrum, the relative intensity is not only 
helpful for assignments, but it is also an important tool to judge if a line may be blended. 
Obviously blended lines were in most cases not included in the line list except for unresolved 
asymmetry doublets and occasionally accidental blends of lines assignable to the ground 
vibrational state of 3-hydroxypropenal as long as the line shape was sufficiently symmetric.

Initial assignments were made above 600~GHz because this is the region in which the $a$-type 
rotation-tunneling transitions occur which were reported by \citet{3-HP_tunneling_1999}. 
However, the transitions easily assignable in the first round were $b$-type $R$-branch transitions 
with $31 \le J \le 50$ and mostly fairly high values of $K_a$, up to 32. The lowest $K_a$ values 
by far in this round were those of a $K_a = 12 - 9$ transition.


\begin{figure}
\centering
\includegraphics[width=9cm,angle=0]{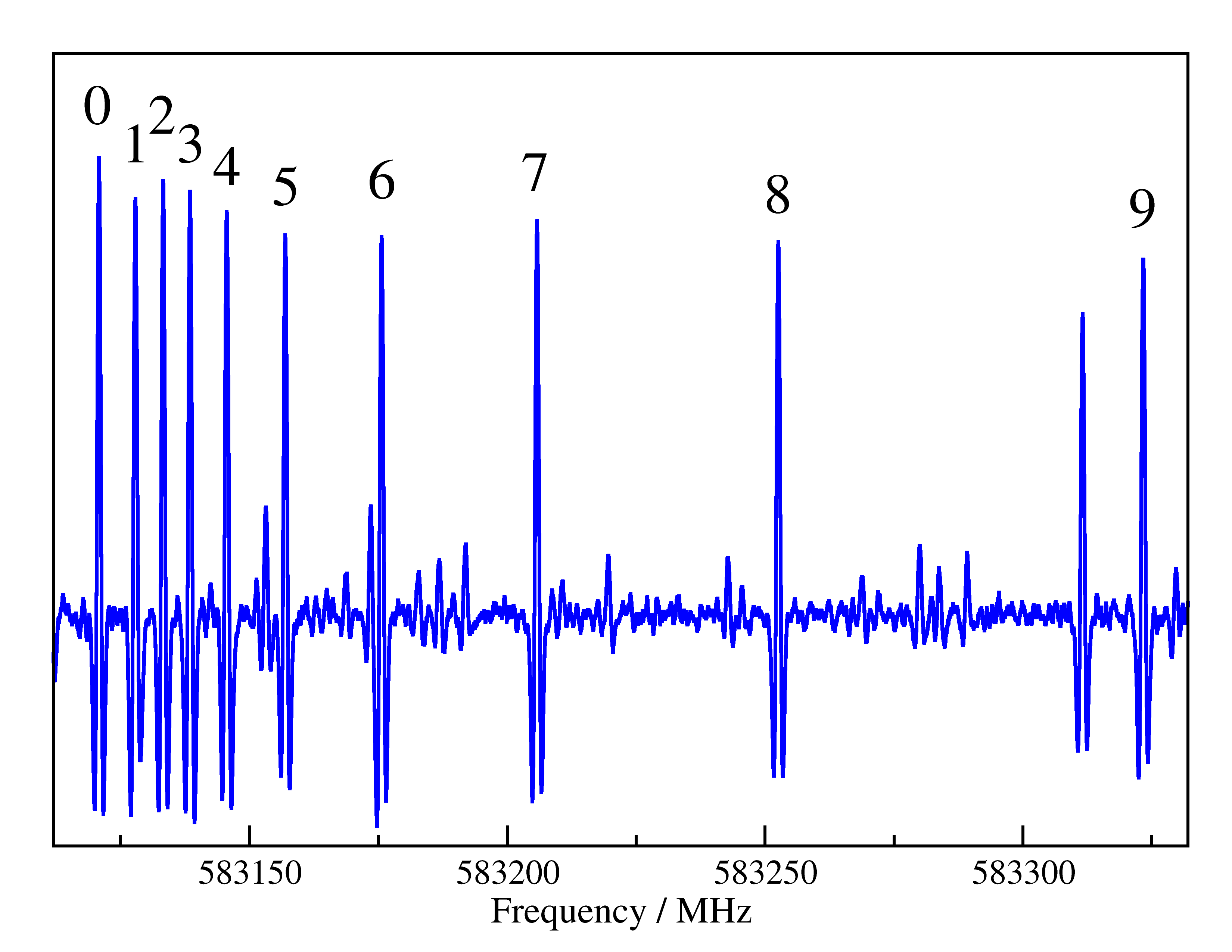}

\caption{Section of the rotational spectrum of 3-hydroxypropenal. The transitions are within 
   $\varv = 0^-$ and have $J_{K_c} = 86$--$n_{86-2n} \gets 85$--$n_{85-2n}$; the values of $n$ 
   are indicated. Transitions with $K_a = 1 - 0$ and $0 - 1$ are paired for $n = 0$, those with 
   $K_a = 2 - 1$ and $1 - 2$ are paired for $n = 1$, and so on.}
\label{oblate_583GHz}
\end{figure}


Not all of the assigned transitions could be fit satisfactorily in the first few rounds. 
Transitions with large residuals were weighted out temporarily as we assumed these residuals 
were a consequence of correlation among the interaction parameters and between these and the 
remaining parameters. Eventually, most of these lines were fit well, the remaining ones were 
omitted as for most of these closer inspection suggested the lines to be blended. 
Transitions with modest residuals around three to four times of the experimental uncertainties 
were marked for occasional inspection. Almost all of these were fit well in later fits.

Subsequently, assignments were made between 500 and 660~GHz in several rounds, slowly increasing 
the quantum number range and also assigning weaker lines. The second round included $a$-type 
rotation-tunneling transitions, and soon thereafter, we assigned transitions with low values 
of $K_a$ at and near the oblate limit, i.e., with $K_c$ near $J$ and $J$ around 80 and 90. 
A conspicuous clustering of $R$-branch transitions occurs near the oblate limit, as shown 
in Fig.~\ref{oblate_583GHz}: transition decreasing in $J$ by one, in $K_c$ by two, and 
increasing in $K_a$ by one occur very close in frequency, the appearance resembling the origin 
of a $Q$-branch.

\begin{figure}
\centering
\includegraphics[width=9cm,angle=0]{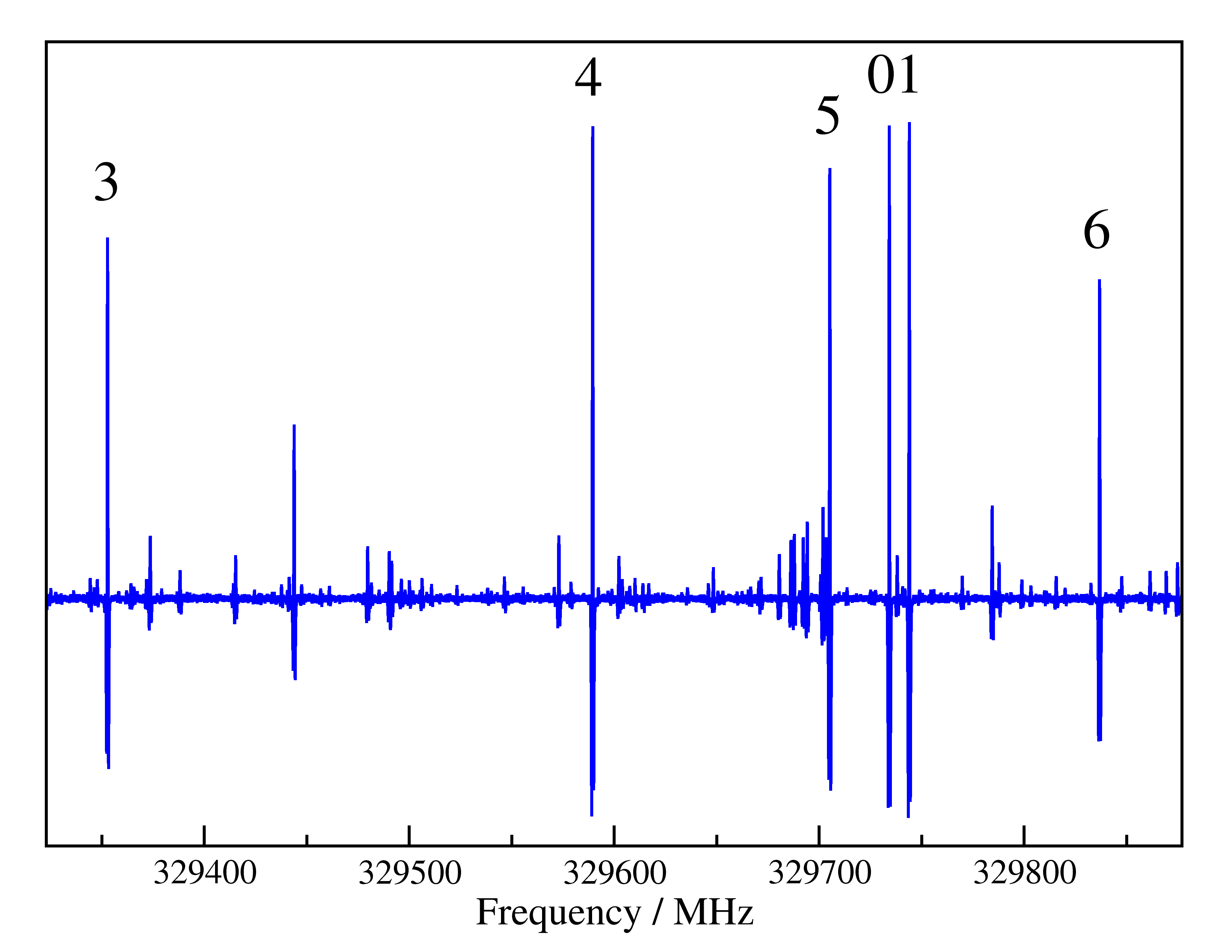}

\caption{Section of the rotational spectrum of 3-hydroxypropenal. The transitions are within 
   $\varv = 0^+$ and have $J_{K_c} = 48$--$n_{48-2n} \gets 47$--$n_{47-2n}$; the values of $n$ are 
   indicated. The $J_{K_c} = 46_{44} - 45_{43}$ transition ($n = 2$) is shifted by $\sim$2~GHz 
   to higher frequencies.}
\label{oblate_330GHz}
\end{figure}

We assigned transitions in sections decreasing in frequency afterwards. The clustering of 
$R$-branch transitions near the oblate limit was a persistent feature, but frequently disrupted 
for some transitions, as displayed in Fig.~\ref{oblate_330GHz}. The disruption is caused by a 
$\Delta K_c = 2$ interaction which is resonant at $J = 45$ and $K_c = 43$ in $\varv = 0^+$ 
and $K_c = 45$ in $\varv = 0^-$. The $\varv = 0^+$ transition with $J_{K_c} = 46_{44} - 45_{43}$ 
is shifted by $\sim$2~GHz to 331903.2~MHz, in the 330 to 400~GHz measurement gap.


\begin{figure}
\centering
\includegraphics[width=9cm,angle=0]{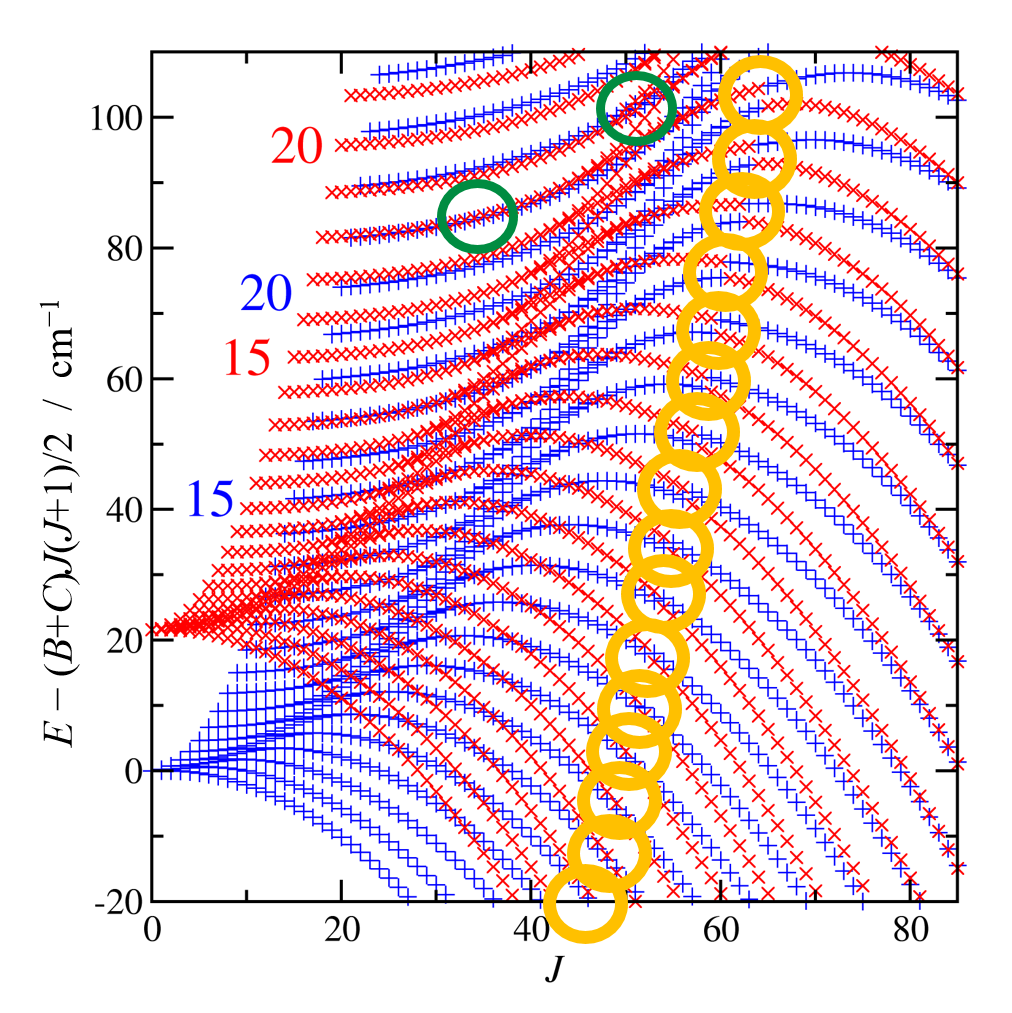}

\caption{Section of the reduced energy plot of 3-hydroxypropenal. Data points of $\varv = 0^+$ 
   are shown in blue, those of $\varv = 0^-$ in red; selected $K_a$ series are indicated to 
   the left. Resonant interactions with $\Delta K_c = 2$ are indicated by yellow circles. 
   They occur at $K_c = 43$ for $\varv = 0^+$ and $K_c = 45$ for $\varv = 0^-$ for 
   $45 \le J \le 48$. The resonant $K_c$ pair increases gradually with $J$, $K_c = 44$ and 
   46 are resonant for $49 \le J \le 58$. Two $\Delta K_a = 3$ interactions are resonant at 
   $K_a = 21/18$ and $J = 35$ as well as $K_a = 22/19$ and $J = 49$ for $ortho$ and $J = 50$ 
   for $para$, which is signaled by green circles. Further $\Delta K_a = 3$ or 5 interactions 
   are usually more local and are not highlighted.}
\label{red_egy}
\end{figure}


A plot of the reduced energy $E - (B + C)J(J + 1)/2$ versus $J$ is shown in Fig.~\ref{red_egy}. 
The purely $J$-depended effects are subtracted off to first order from the rotational energy 
such that the combined effects of asymmetry, centrifugal distortion, and perturbation are visible 
for individual $K_a$ series of $\varv = 0^+$ and $\varv = 0^-$. The $\Delta K_c = 2$ interaction 
is resonant at $K_c = 43$ for $\varv = 0^+$ and $K_c = 45$ for $\varv = 0^-$ for $45 \le J \le 48$. 
The resonant $K_c$ pair increases gradually with $J$, $K_c = 44$ and 46 are resonant for 
$49 \le J \le 58$, $K_c = 45$ and 47 for $59 \le J \le 63$, $K_c = 46$ and 48 for $64 \le J \le 66$, 
and so on. The perturbations are fairly weak and rather local for the lower values of $J$, 
the shift in energy is not or only hardly seen in the reduced energy plot. Increasingly larger 
effects are noticeable in Fig.~\ref{red_egy} for $J > 50$. At least two, often many more transitions 
were included in the line list for each tunneling state and nearly all $J$ between 45 and 70, 
at which point many energy levels with similar quantum numbers are perturbed very strongly, 
and the attribution of energy levels to $\Delta K_c = 2$ interactions are complex.

We identified other strong interactions which cause perturbations in several values of $J$, 
for example, with $\Delta K_a = 3$, which are resonant at $J = 35$ and $K_a = 21$ and 18 for 
$\varv = 0^+$ and $\varv = 0^-$, respectively, and at $J = 49$ and 50 for $ortho$- and 
$para$-3-hydroxypropenal, respectively, in the case of $K_a = 22$ and 19. These resonances are 
also indicated in Fig.~\ref{red_egy}. Multiple transitions were assigned for these levels as well 
as in the vicinity of the resonance for both tunneling states. Other, more local perturbations 
are best described as connecting levels with $\Delta K_c = 4$, $\Delta K_a$ changes by 5 for one 
asymmetry side and by 3 for the other. Two examples with multiple transitions in the line list 
are $37_{17,21}$ in $\varv = 0^+$ and $37_{12,25}$ in $\varv = 0^-$ or the respective pair 
$41_{16,26}$ and $41_{11,30}$. The various perturbations may transfer intensity to 
rotation-tunneling transitions having $\Delta K_a = 2$, 4, and 6.

\begin{figure}
\centering
\includegraphics[width=9cm,angle=0]{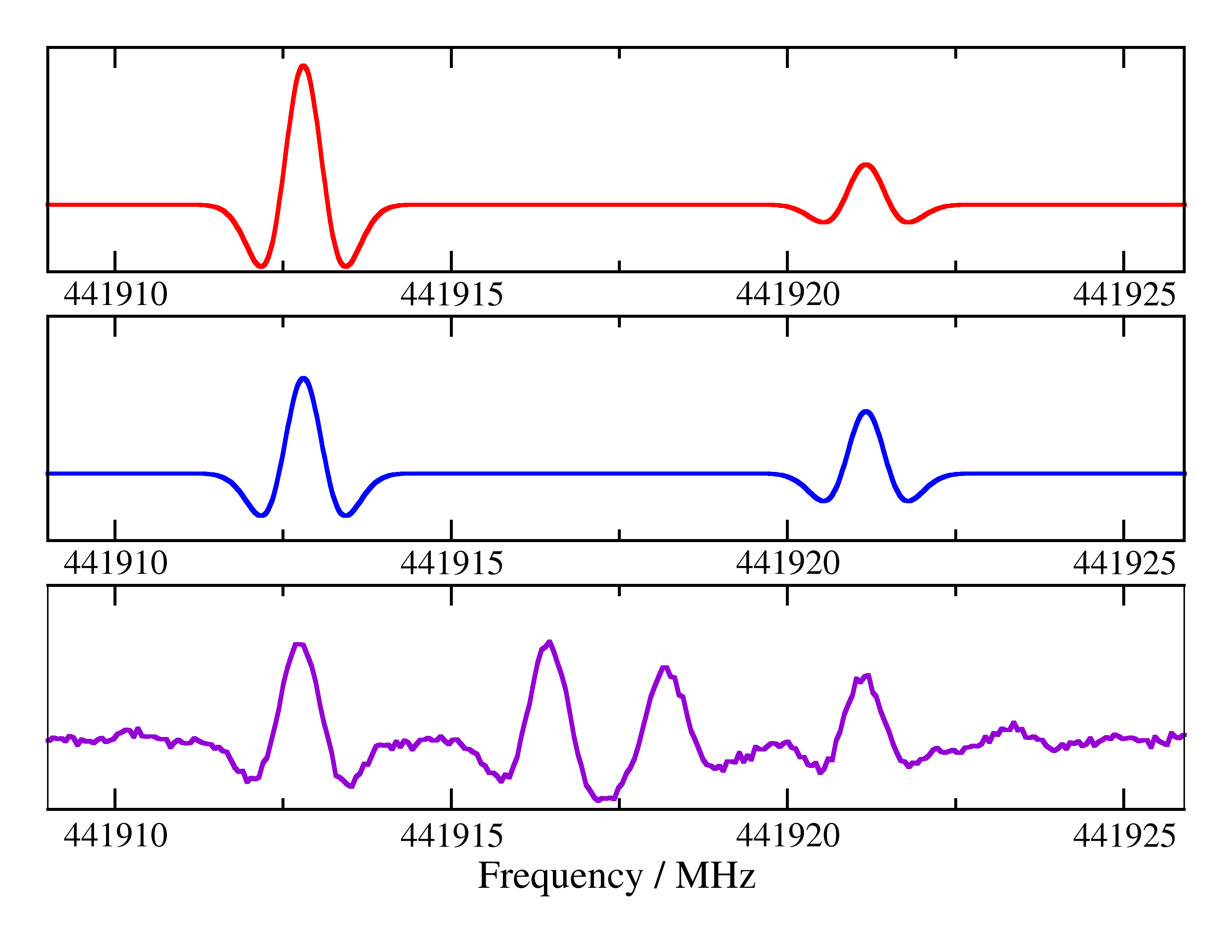}

\caption{Section of the rotational spectrum of 3-hydroxypropenal in the lower trace. 
   The $J_{K_a} = 58_{41} - 58_{40}$ transition within $\varv = 0^-$ is shown to the left, and the 
   $42_{41} - 42_{40}$ transition within $\varv = 0^+$ to the right. The two transition between 
   them are unassigned. The upper trace shows a simulation of the spectrum with $\mu _a$, $\mu _b$, 
   and $F_{ab}$ chosen to be positive; the sign of $\mu _a$ is negative in the middle trace.}
\label{sign-mu_a}
\end{figure}

After having reached the lower frequency limit of 150~GHz in the assignment process, further 
assignments were made by going up in frequency to the upper limit. The assignments involved 
often weak lines, but also some stronger lines whose assignments were uncertain initially because 
they were calculated too far away from their frequency in the experimental spectrum. In addition, 
few transitions were checked that had relatively large residuals in the fit. 
Among the weak transitions included in the line list are $b$-type $Q$-branch rotational transitions 
with $K_a$ up to 59. Some of these displayed a fairly strong dependence on the sign choice of 
$\mu _a$ with $\mu _b$ and $F_{ab}$ chosen to be positive, as can be seen in Fig.~\ref{sign-mu_a}.
The upper trace simulation shows two lines differing in intensity by a factor of $\sim$3.5 whereas 
in the middle trace simulation the factor is only $\sim$1.5, which is clearly more appropriate. 
These intensity changes upon sign change of $\mu _a$ were among the largest easily discernible 
in the spectrum.


\begin{table*}
\begin{center}
\caption{Average spectroscopic parameters $X$ (MHz) and energy $E$ with rotational corrections or changes $\Delta X$ from 
         average spectroscopic parameters of 3-hydroxypropenal along with interaction parameters determined in the present study.}
\label{spec-parameters}
\renewcommand{\arraystretch}{1.10}
\begin{tabular}[t]{lr@{}lcr@{}ll}
\hline \hline
Parameter~I            & \multicolumn{2}{c}{Value~I} & & \multicolumn{2}{c}{Value~II} & Parameter~II                           \\
\hline
                            &        &               & &  323523&.09353~(185)   & $E$ / $\Delta E^a$                           \\
                            &        &               & &      24&.8460491~(130) & $E_K$$^b$                                    \\
                            &        &               & &   $-$18&.7394637~(47)  & $E_J$$^b$                                    \\
                            &        &               & &    $-$4&.0806764~(23)  & $E_2$$^b$                                    \\
$A$                         &    9839&.9556351~(204) & &       6&.1065853~(134) & $\Delta A$$^c$                               \\
$B$                         &    5185&.6270046~(137) & &   $-$26&.9008165~(77)  & $\Delta B$$^c$                               \\
$C$                         &    3393&.8067801~(96)  & &   $-$10&.5781109~(51)  & $\Delta C$$^c$                               \\
$D_K \times 10^3$           &       4&.744792~(60)   & &       0&.32757~(108)   & $(E_{KK}   / -\Delta D_K)    \times 10^3$    \\
$D_{JK} \times 10^3$        &    $-$3&.449702~(41)   & &    $-$0&.79152~(112)   & $(E_{JK}   / -\Delta D_{JK}) \times 10^3$    \\
$D_J \times 10^3$           &       3&.1237538~(65)  & &       0&.455189~(37)   & $(E_{JJ}   / -\Delta D_J)    \times 10^3$    \\
$d_1 \times 10^3$           &    $-$1&.2433617~(32)  & &       0&.1810947~(22)  & $(E_{2J}   / \Delta d_1)     \times 10^3$    \\
$d_2 \times 10^3$           &    $-$0&.1596366~(14)  & &       0&.018008~(18)   & $(E_4      / \Delta d_2)     \times 10^3$    \\
$H_K \times 10^9$           &      46&.4796~(265)    & &   $-$18&.450~(105)     & $(E_{KKK}  / \Delta H_K)     \times 10^9$    \\
$H_{KJ} \times 10^9$        &   $-$73&.3040~(243)    & &       6&.817~(140)     & $(E_{JKK}  / \Delta H_{KJ})  \times 10^9$    \\
$H_{JK} \times 10^9$        &      39&.0206~(109)    & &      15&.617~(41)      & $(E_{JJK}  / \Delta H_{JK})  \times 10^9$    \\
$H_J \times 10^9$           &    $-$5&.34499~(167)   & &    $-$4&.29228~(125)   & $(E_{JJJ}  / \Delta H_J)     \times 10^9$    \\
$h_1 \times 10^9$           &    $-$1&.99143~(83)    & &    $-$2&.18286~(75)    & $(E_{2JJ}  / \Delta h_1)     \times 10^9$    \\
$h_2 \times 10^9$           &       0&.354894~(173)  & &    $-$0&.35670~(38)    & $(E_{4J}   / \Delta h_2)     \times 10^9$    \\
$h_3 \times 10^9$           &       0&.097037~(72)   & &    $-$0&.065374~(202)  & $(E_6      / \Delta h_3)     \times 10^9$    \\
$L_K \times 10^{15}$        &  $-$300&.7~(87)        & &     763&.3~(172)       & $(E_{KKKK} / \Delta L_K)     \times 10^{15}$ \\
$L_{KKJ} \times 10^{15}$    &     416&.7~(109)       & & $-$1124&.1~(250)       & $(E_{JKKK} / \Delta L_{KKJ}) \times 10^{15}$ \\
$L_{JK} \times 10^{15}$     &   $-$45&.9~(58)        & &     685&.3~(134)       & $(E_{JJKK} / \Delta L_{JK})  \times 10^{15}$ \\
$L_{JJK} \times 10^{15}$    &  $-$108&.45~(169)      & &  $-$325&.9~(32)        & $(E_{JJJK} / \Delta L_{JJK}) \times 10^{15}$ \\
$L_J \times 10^{15}$        &      19&.254~(158)     & &   $-$23&.883~(163)     & $(E_{JJJJ} / \Delta L_J)     \times 10^{15}$ \\
$l_1 \times 10^{15}$        &       9&.019~(79)      & &   $-$21&.516~(84)      & $(E_{2JJJ} / \Delta l_1)     \times 10^{15}$ \\
$l_2 \times 10^{15}$        &        &               & &   $-$11&.114~(59)      & $(E_{4JJ}  / \Delta l_2)     \times 10^{15}$ \\
$l_3 \times 10^{15}$        &       0&.3182~(118)    & &        &               & $(E_{6J}   / \Delta l_3)     \times 10^{15}$ \\
$l_4 \times 10^{15}$        &    $-$0&.1252~(34)     & &       0&.1665~(34)     & $(E_8      / \Delta l_4)     \times 10^{15}$ \\ 
$F_{ab}$                    &      45&.3158~(32)     & &        &               &                                              \\
$F_{ab,K} \times 10^6$      &  $-$728&.03~(70)       & &        &               &                                              \\
$F_{ab,J} \times 10^6$      &     906&.35~(37)       & &        &               &                                              \\
$F_{2ab} \times 10^6$       &      53&.629~(249)     & &        &               &                                              \\
$F_{ab,KK} \times 10^9$     &   $-$11&.738~(74)      & &        &               &                                              \\
$F_{ab,JK} \times 10^9$     &      27&.0669~(272)    & &        &               &                                              \\
$F_{2ab,J} \times 10^9$     &       7&.9407~(224)    & &        &               &                                              \\
$F_{ab,JKK} \times 10^{12}$ &       0&.1050~(91)     & &        &               &                                              \\
$F_{ab,JJK} \times 10^{12}$ &    $-$0&.9419~(40)     & &        &               &                                              \\
\hline
\end{tabular}
\end{center}
\tablefoot{
Watson's $S$ reductions was used in the representation $I^r$. Numbers in parentheses are one standard 
deviation in units of the least significant figures. All parameters are defined positively, except 
$D_K$, $D_{JK}$, $D_J$, and their differences. Empty fields indicate parameters not used in the final fit. 
Parameters $X$ and interaction parameters $F_{ab}$ etc. are listed under headings Parameter~I and Value~I; 
the $\Delta X$, which can be interpreted as rotational corrections to $E$, are given under the headings 
Value~II and Parameter~II. 
$^{(a)}$ $E(0^-) - E(0^+) = 2E = 2\Delta E = 647046.1871~(37)$~MHz or 21.58313759~(12)~cm$^{-1}$ 
per definition of $E$ and $\Delta E$, see section~\ref{analysis}. 
$^{(b)}$ Applied in the fit with rotational corrections to the energies. 
$^{(c)}$ Applied in the fit with differences in spectroscopic parameters. 
} 
\end{table*}


Our final line list consisted of 11551 transitions from the present study which correspond 
to 7551 different frequencies, mostly because of unresolved asymmetry splitting. The quantum 
numbers $J$, $K_a$, and $K_c$ reach values of 100, 59, and 97. The numbers of transitions 
within $\varv = 0^+$ and $\varv = 0^-$ is 5123 and 5127, respectively, most having 
$\Delta K_a = \pm1$, some have $\Delta K_a = 3$. There are 1066 transitions from 
$\varv = 0^+$ to $\varv = 0^-$, mostly with $\Delta K_a = 0$, some with $\Delta K_a = -2$, 
and even fewer with $\Delta K_a = -4$ or $-6$. And finally there are 235 transitions from 
$\varv = 0^-$ to $\varv = 0^+$, mostly with $\Delta K_a = +4$ and +2, but also 16 having 
$\Delta K_a = +6$. The line list also contains 79 lines from \citet{3-HP_rot_etc_1981} 
and 11 lines from \citet{3-HP_rot_Zeeman_1983}, see also Section~\ref{prev-data}.

The final set of spectroscopic parameters is presented in Table~\ref{spec-parameters}. 
The rms of our lines is 17.4~kHz, and the rms error is 0.838. The rms of the previous 
data from \citet{3-HP_rot_etc_1981} and from \citet{3-HP_rot_Zeeman_1983} is 8.9~kHz 
and 7.4 kHz, respectively. The rms of 17.3~kHz and the rms error of 0.839 for the total fit 
are obviously dominated by our data.

The line, parameter, and fit file, along with auxiliary files, are available in the fitting 
spectra section\footnote{https://cdms.astro.uni-koeln.de/classic/predictions/pickett/beispiele/3-HP/} 
of the CDMS. A calculation of the rotation-tunneling spectrum is deposited in the catalog 
section\footnote{https://cdms.astro.uni-koeln.de/classic/entries/} of the CDMS. 
The line list with quantum numbers, uncertainties and residuals between observed frequencies 
and those calculated from the final set of spectroscopic parameters is available from ads.


\begin{figure}
\centering
\includegraphics[width=9cm,angle=0]{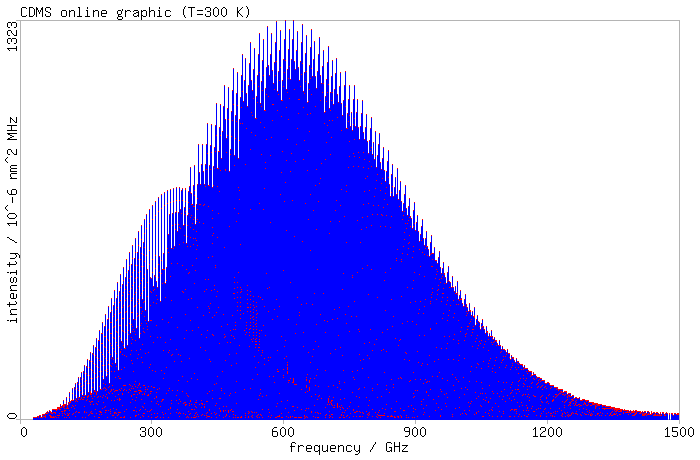}\\
\includegraphics[width=9cm,angle=0]{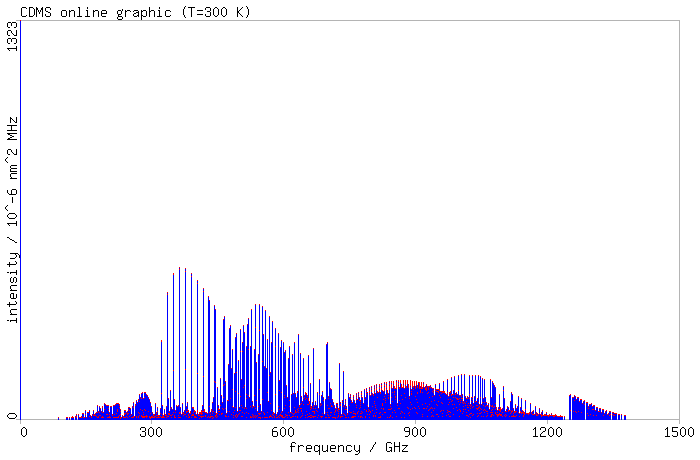}

\caption{Stick spectra of 3-hydroxypropenal. The upper trace shows the rotational transitions 
   with $b$-type selection rules, and the lower trace displays the rotation-tunneling 
   transitions with $a$-type selection rules.}
\label{stick-spectrum}
\end{figure}

\section{Discussion of the laboratory spectroscopic results}
\label{lab-discussion}

It is somewhat easier to compare the differences in spectroscopic parameters with the respective 
spectroscopic parameters. The lowest order parameters $\Delta A$, $\Delta B$, and $\Delta C$ are 
comparatively large, though still much smaller than $A$, $B$, and $C$ because the tunneling 
barrier is relatively low, and the splitting between $\varv = 0^+$ and $\varv = 0^-$ is relatively 
large. It is very common that the ratios of the differences in spectroscopic parameters with 
the respective spectroscpic parameters increase with order of magnitude of the parameters. 
The differences are only about an order of magnitude smaller in case of the quartics, 
similar in magnitude for the sextics, and mostly slightly larger in case of the octics. 
The number of transitions within each tunneling state are much larger than those between the 
tunneling states. Therefore, it is not surprising that the absolute uncertainties of the changes 
in the distortion parameters are mostly larger than those of the respective parameters; however, 
it is opposite in the case of the rotational parameters.

The spectroscopic parameters from the present work are accurate enough for all types of radio 
observations. A calculation at 300~K up to 1.5~THz contains only very few lines with calculated 
uncertainties exceeding 0.1~MHz; these very few lines should probably be viewed with some caution. 
This is very different to the line list prior to this work. 
A calculated spectrum at 125~K contained several fairly strong transitions with uncertainties 
of around 10~MHz in the range of the PILS data; these transitions are only about a factor of 5 
weaker than the strongest transitions in this range. These are $a$-type transitions with low values 
of $K_a$ and $\Delta K_a = 2$; their presently calculated frequencies differ by about three times 
the initial uncertainties. Several other fairly strong transitions had uncertainties around 1~MHz 
or more, and deviations of up three times these uncertainties are quite common.
Stick spectra of the rotational and rotation-tunneling transitions are shown in Fig.~\ref{stick-spectrum}. 
While the rotation-tunneling transitions are usually much weaker, relatively strong transitions appear 
in the 300 to 700~GHz region.

The rotation of the RAS with respect to the principal inertial axis system is derived as 
$\sin2\theta = 2F_{ab}/(A - B)$ \citep{RAS_Pickett_1972}. Its value of 0.557884 (39)$^{\rm o}$ 
appears to be a fairly typical one. Data derived for selected other molecules are, for example, 
0.8390$^{\rm o}$ ($ab$-plane) and 1.7892$^{\rm o}$ ($bc$) for $aGg'$-ethylene glycol 
\citep{aGg-eglyc_rot_2003}, as well as 0.073$^{\rm o}$ ($ac$) and 0.286$^{\rm o}$ ($bc$) 
for propargyl alcohol \citep{PrgOH_rot_2005}.

A comparison of our present spectroscopic parameters with those from previous work is meaningful for 
low order parameters. The rotation-tunneling interaction was considered by \citet{3-HP_analysis_2_1984}, 
by \citet{3-HP_FIR_1991}, and by \citet{3-HP_tunneling_1999}. 
Rotational and quartic centrifugal distortion parameters, the energy difference, the interaction 
parameter $F_{ab}$, and its distortion correction $F_{ab,J}$ were determined in all three studies. 
The agreement among the rotational parameters is good, while the agreement is only good for the 
quartic parameters from the latest study \citep{3-HP_tunneling_1999} and reasonable in the case 
of the other two studies. It is instructive to compare the energy differences and interaction 
parameters from previous work with ours. The energy differences are 647049 (12) MHz 
\citep{3-HP_analysis_2_1984}, 647094.9 (51) MHz \citep{3-HP_FIR_1991}, and 647046.208 (19) MHz 
\citep{3-HP_tunneling_1999}, compared to 647046.1871 (37) MHz from our study. The agreement is 
excellent in the case of the latest study and reasonable for the earlier ones; the smaller 
uncertainties reflect to some degree the extent of the corresponding data set. 
The interaction parameters in the same order are 45.51 (4) MHz, 46.01 (2) MHz, and 45.8965 (82) MHz, 
compared to 45.3158 (32) MHz from our study. The agreement is fairly good in all cases, the somewhat 
large deviation from the latest study with respect to the combined uncertainties is probably caused 
by a much larger set of parameters needed to fit our very large data set. 
\citet{3-HP_FIR_1991} also reported a nearly full set of sextic distortion parameters while 
\citet{3-HP_tunneling_1999} reported a set of four diagonal distortion parameters for each 
tunneling state plus a value for $F_{ab,K}$; the agreement with our values is reasonable.

\citet{3-HP_analysis_1_1984} evaluated the tunneling barrier height for the main isotopolog as 
$\sim$6.6~kcal/mol or $\sim$2300~cm$^{-1}$. While the authors pointed out that the small amount 
of data for the main isotopic species leads to a great uncertainty of this barrier height, we 
note that we are not aware of later attempts to determine the barrier height in similar ways with 
more data, possibly because the methods were deemed to be too simple. The $\sim$6.6~kcal/mol may 
be compared with 4.1~kcal/mol or about 1400~cm$^{-1}$ derived from a theoretical potential energy 
surface (PES) \citep{3-HP_PES_2008}.

The accurate derivation of tunneling splittings from theoretical PESs is challenging. 
\citet{3-HP_PES_2008} present ground state splittings of 21.6 and 22.6~cm$^{-1}$ from calculations 
in Cartesian and normal coordinates, respectively, with estimated uncertaities of 2$-$3~cm$^{-1}$. 
Both values compare very favorably with our present 21.58313759 (12) cm$^{-1}$, which in turn is 
very similar to most earlier, less accurate values, as discussed further above. 
\citet{3-HP_exp_tunneling_exc-st_2013} determined experimentally tunneling splittings in excited 
vibrational states of 3-hydroxypropenal and compare the values with those from several theoretical 
calculations. While the agreement is in all but one instances very good for the ground state splitting, 
the agreement is less favorable in excited vibrational states.


\begin{figure*}
\centering
\includegraphics[width=1.0\linewidth]{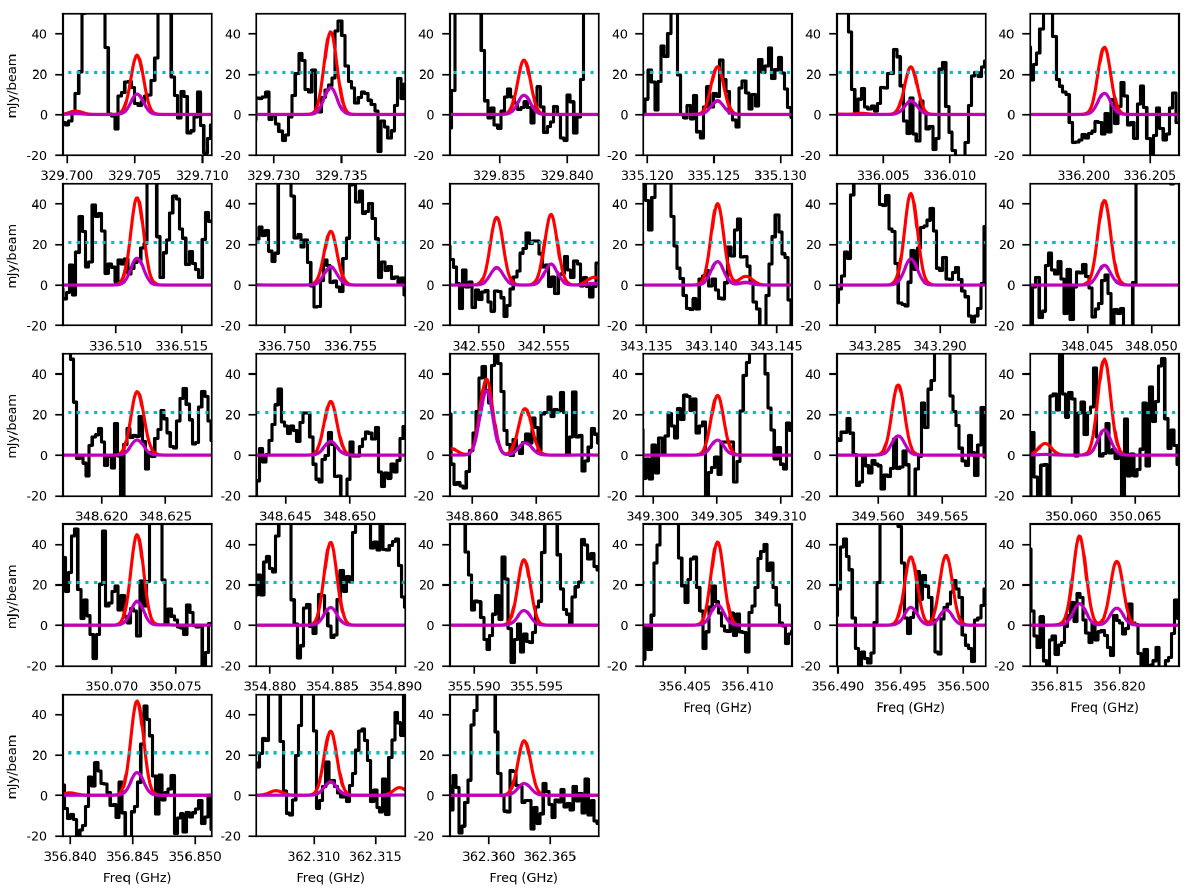}
\caption{\label{fig_undetect} Undetected lines of 3-hydroxypropenal that are predicted above 
3$\sigma$ with the new spectroscopic data and the parameters previously constrained by 
\citet{3-HP_det_2022} for an excitation temperature $T_{\rm ex}$ = 300~K. The PILS observations 
are in black. The model in red has been obtained with the best-fit parameters derived by 
\citet{3-HP_det_2022} for $T_{\rm ex}$ = 300~K, i.e. with a column density 
$N = 1.8 \times 10^{15}$~cm$^{-2}$. For comparison, the model obtained for $T_{\rm ex} = 125$~K 
($N = 1.0 \times 10^{15}$~cm$^{-2}$) is added in magenta. The cyan dotted line represents the 
3~rms level per 0.33 km\,s$^{-1}$ (1/3 of the line width).}
\end{figure*}


\section{Observations of 3-hydroxypropenal in IRAS16293~B}
\label{ALMA-observations}
\subsection{Analysis of the ALMA-PILS observations}
\label{PILS-observations}

The 3-hydroxypropenal molecule has been tentatively detected in the solar-type protostar 
IRAS16293~B by \citet{3-HP_det_2022}. The lines have been searched for in PILS 
carried out with ALMA. This survey covers a large spectral range between 329.1 and 362.9~GHz 
with an angular resolution of 0.5$\arcsec$ ($\sim$60~au) and a spectral resolution of 0.244~MHz 
($\sim$0.2 km\,s$^{-1}$). The data reduction process as well as additional characteristics 
of the observations are presented in \citet{PILS_overview_2016}. 
Thanks to its high sensitivity (rms $\approx$ 4-5~mJy\,beam$^{-1}$ per km\,s$^{-1}$), a large variety 
of complex organic molecules have been detected in this survey. Local thermodynamic equilibrium 
(LTE) models have been carried out to reproduce the observations and derive the column densities 
and excitation temperatures of the different molecules. In particular, different excitation 
temperatures have been found in this source. Some molecules, such as NH$_2$CHO, NH$_2$CN, CH$_3$OH, 
HCOOH, CH$_3$OCHO, CH$_2$OHCHO, and CH$_3$CH$_2$OH show a high excitation temperature of 
300~K \citep{deuterated_PILS_2016,H2NCN_PILS_2018,PILS_div-isos_2018}, while others 
($c$-C$_2$H$_4$O, CH$_3$CHO, CH$_3$OCH$_3$, CH$_3$CCH, CH$_3$CN, C$_2$H$_5$CN) show a lower excitation 
temperature of 125~K \citep{PILS_COMs_2017,PILS_div-isos_2018,CH3CN_PILS_2018,CH3CCH_PILS_2019}. 
Based on the previous catalog of 3-hydroxypropenal lines, it was not possible to distinguish 
between an excitation temperature of 125~K and 300~K for this species. Indeed, the lines with 
high $E_{\rm up}$ that should allow us to constrain the excitation temperature, showed too large 
frequency uncertainties. Given the very high line density of this survey, it was not possible 
to know if they were detected. Two column densities were consequently derived: 
$N = 1.0 \times 10^{15}$~cm$^{-2}$ for $T_{\rm ex} = 125$~K and $N = 1.8 \times 10^{15}$~cm$^{-2}$ 
for $T_{\rm ex}$ = 300~K.

The results of this new spectroscopic study enable us to search for the lines of 3-hydroxypropenal 
of high excitation. We ran two LTE models with the previously derived best-fit parameters and 
compared them to the observations. A similar result is obtained for the 11 lines attributed to 
3-hydroxypropenal, which showed moderate $E_{\rm up}$ values (< 300~K) and small frequency 
uncertainties (see Fig.~2 of \citealt{3-HP_det_2022}). If we check the full survey, no major issue 
is seen for the model with $T_{\rm ex}$ = 125~K. However, the model with $T_{\rm ex}$ = 300~K and 
$N = 1.8 \times 10^{15}$~cm$^{-2}$ does not reproduce the observations. Indeed, several undetected 
lines ($\sim$30) are predicted above 3$\sigma$ with this model (see Fig.~\ref{fig_undetect}). 
A better agreement would be obtained if we lower the column density at $1.0 \times 10^{15}$~cm$^{-2}$. 
But several detected lines (Fig.~2 of \citealt{3-HP_det_2022}) are in that case under-produced 
in terms of intensity. In summary, an excitation temperature of 125~K is more appropriate than 300~K 
for this species.

\begin{figure*}
\centering
\includegraphics[width=1.0\linewidth]{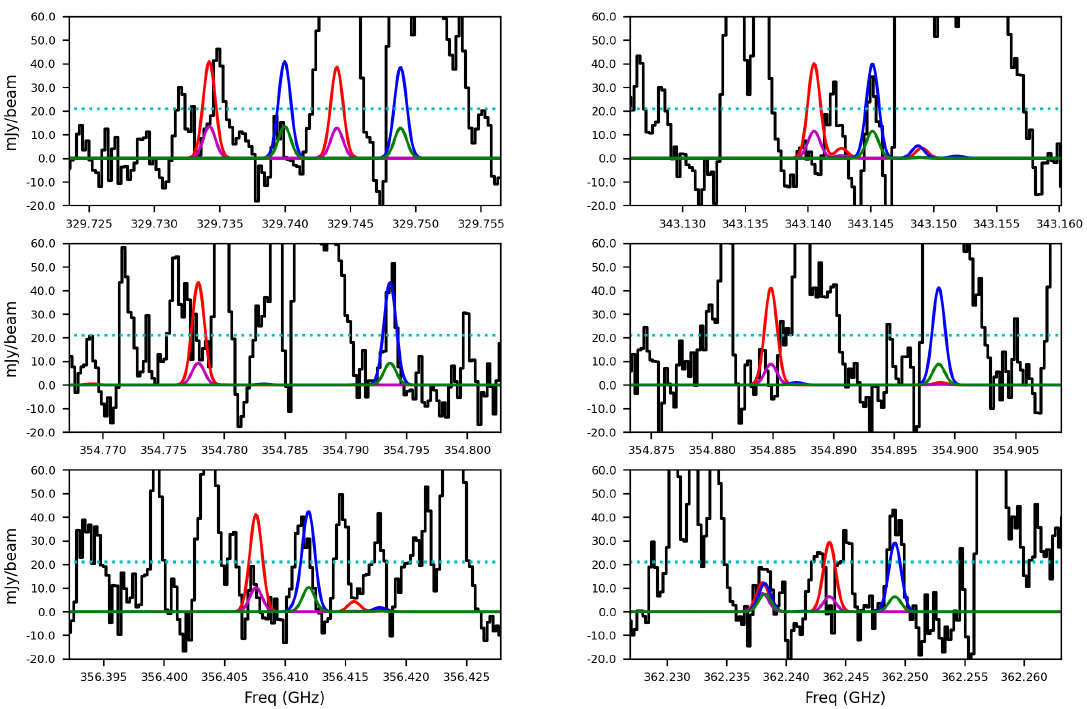}
\caption{\label{fig_shift} Examples of high excitation transitions of 3-hydroxypropenal that 
have been significantly shifted with the new spectroscopic study. Models obtained with previous 
spectroscopy are indicated in blue ($T_{\rm ex}$ = 300~K) and green ($T_{\rm ex} = 125$~K). 
The new models are in red ($T_{\rm ex} = 300$~K) and magenta ($T_{\rm ex} = 125$~K). 
The column densities are $N = 1.0 \times 10^{15}$~cm$^{-2}$ for $T_{\rm ex} = 125$~K and 
$N = 1.8 \times 10^{15}$~cm$^{-2}$ for $T_{\rm ex} = 300$~K. An absorption due to a nearby 
bright line is present around 343.14~GHz. The cyan dotted line represents the 3~rms level 
per 0.33 km\,s$^{-1}$ (1/3 of the line width).}
\end{figure*}

Even if we do not detect any high excitation lines of 3-hydroxypropenal in the PILS data, 
it is important to note that such a spectroscopic study is useful, not only to constrain 
the excitation temperature of the molecule in IRAS16293~B but also for its search in other sources 
with potentially higher excitation temperatures. The shift of some of the high excitation lines 
of 3-hydroxypropenal is significant. For example, in Figure~\ref{fig_shift}, we show that previous 
high excitation transitions that were matching some line emission in IRAS16293~B are now predicted 
in areas free of lines. A proper line identification of this molecule in other sources is now possible.


\subsection{Search for additional transitions of 3-hydroxypropenal in the ALMA archive}
\label{ALMA-archive}

We searched the ALMA archive for projects possibly covering additional transitions 
to further strengthen the assignment of 3\nobreakdash-hydroxypropenal toward IRAS16293~B. 
We used the ATOMIS\footnote{ALMA archive TOol for Molecular Investigations in Space, 
\url{https://atomis.irap.omp.eu/}} web application \citep{Coutens2023} to list 
all the ALMA archival observations that cover transitions of 3\nobreakdash-hydroxypropenal 
with $A_{\rm ij} \geq 5 \times 10^{-4}$~s$^{-1}$ and $E_{\rm up} \leq 500$~K. 
The angular resolution was limited to the range 0.3--0.7$\arcsec$ so that it is 
relatively close to the PILS resolution ($\sim$0.5$\arcsec$). We requested data 
with a sensitivity better than 10~mJy\,beam$^{-1}$ for a bin width of 1~km\,s$^{-1}$ and 
only selected observational results with a spectral resolution better than 0.5~km\,s$^{-1}$.
Table~\ref{tab_archive} summarizes the observational projects that cover such transitions 
(apart from the PILS project).

\begin{table*}[h!]
\caption{Observations covering transitions of 3-hydroxypropenal in the ALMA archive.}
\begin{center}
\begin{tabular}{@{}crr@{~~}c@{~}ccl}
\hline
\hline
Transition & Frequency & $E_{\rm up}$ & $A_{\rm ij}$ & Project & Ang. res. & Comments \\
$J',K_a',K_c' (\varv _t') - J'',K_a'',K_c'' (\varv _t'')$ & \multicolumn{1}{c}{(GHz)} & (K) & 
($\times$\,10$^{-4}$ s$^{-1}$) & & ($\arcsec$ $\times$ $\arcsec$) & \\
\hline
35,0,35 ($0^-$) --  34,1,34 ($0^-$)     & 240.06964 & 241.3 & 5.1  & 2021.1.01164.S & 0.71 $\times$ 0.45 & Detected \\
35,1,35 ($0^-$) --  34,0,34 ($0^-$)     & 240.06964 & 241.3 & 5.1  & 2021.1.01164.S & 0.71 $\times$ 0.45 &  ---      \\
34,1,33 ($0^-$) --  33,2,32 ($0^-$)$^a$ & 240.08569 & 240.5 & 4.8  & 2021.1.01164.S & 0.71 $\times$ 0.45 &  Detected \\
34,2,33 ($0^-$) --  33,1,32 ($0^-$)$^a$ & 240.08569 & 240.5 & 4.8  & 2021.1.01164.S & 0.71 $\times$ 0.45 & ---      \\
14,13,1 ($0^+$) --  13,12,2 ($0^+$)     & 258.93983 & 88.4  & 5.7  & 2017.1.01565.S & 0.45 $\times$ 0.37 & Detected \\
14,13,2 ($0^+$) --  13,12,1 ($0^+$)     & 258.93983 & 88.4  & 5.7  & 2017.1.01565.S & 0.45 $\times$ 0.37 & ---      \\
35,4,32 ($0^-$) --  34,3,31 ($0^-$)     & 260.41436 & 273.4 & 5.4  & 2017.1.01565.S & 0.45 $\times$ 0.37 & Blended with HCOOCH$_3$ \\
35,3,32 ($0^-$) --  34,4,31 ($0^-$)     & 260.41436 & 273.4 & 5.4  & 2017.1.01565.S & 0.45 $\times$ 0.37 & and CHD(OH)CHO \\
34,5,30 ($0^-$) --  33,4,29 ($0^-$)     & 260.49075 & 270.5 & 5.0  & 2017.1.01565.S & 0.45 $\times$ 0.37 & Detected \\
34,4,30 ($0^-$) --  33,5,29 ($0^-$)     & 260.49075 & 270.5 & 5.0  & 2017.1.01565.S & 0.45 $\times$ 0.37 & ---      \\
19,12.8 ($0^+$) --  18,11,7 ($0^+$)$^b$ & 291.42772 & 117.9 & 5.2  & 2017.1.01565.S & 0.41 $\times$ 0.37 & Noise    \\
19,12.7 ($0^+$) --  18,11,8 ($0^+$)$^b$ & 291.42936 & 117.9 & 5.2  & 2017.1.01565.S & 0.41 $\times$ 0.37 & Detected \\
36,7,29 ($0^+$) --  35,8,28 ($0^+$)$^b$ & 296.40275 & 295.1 & 5.9  & 2022.1.00554.S & 0.59 $\times$ 0.47 & Noise    \\
36,8,29 ($0^+$) --  35,7,28 ($0^+$)$^b$ & 296.40508 & 295.1 & 5.9  & 2022.1.00554.S & 0.59 $\times$ 0.47 & Blended with U-line \\
43,2,41 ($0^-$) --  42,3,40 ($0^-$)     & 307.72880 & 372.2 & 9.9  & 2022.1.00554.S & 0.57 $\times$ 0.46 & Detected \\
43,3,41 ($0^-$) --  42,2,40 ($0^-$)     & 307.72880 & 372.2 & 9.9  & 2022.1.00554.S & 0.57 $\times$ 0.46 & ---      \\
42,3,39 ($0^-$) --  41,4,38 ($0^-$)     & 307.73151 & 370.0 & 9.4  & 2022.1.00554.S & 0.57 $\times$ 0.46 & Detected \\
42,4,39 ($0^-$) --  41,3,38 ($0^-$)     & 307.73151 & 370.0 & 9.4  & 2022.1.00554.S & 0.57 $\times$ 0.46 & with nearby absorption line \\
41,4,37 ($0^-$) --  40,5,36 ($0^-$)     & 307.76332 & 370.1 & 8.8  & 2022.1.00554.S & 0.57 $\times$ 0.46 & Blended with $^{13}$CH$_3$C(O)CH$_3$  \\
41,5,37 ($0^-$) --  40,4,36 ($0^-$)     & 307.76332 & 370.1 & 8.8  & 2022.1.00554.S & 0.57 $\times$ 0.46 & or SO$^{18}$O \\
44,1,43 ($0^-$) --  43,2,42 ($0^-$)     & 307.76735 & 373.6 & 10.4 & 2022.1.00554.S & 0.57 $\times$ 0.46 & Blended with CH$_3$CHO  \\
44,2,43 ($0^-$) --  43,1,42 ($0^-$)     & 307.76735 & 373.6 & 10.4 & 2022.1.00554.S & 0.57 $\times$ 0.46 & + absorption \\
37,9,29 ($0^-$) --  36,8,28 ($0^-$)     & 308.76184 & 347.5 & 6.0  & 2022.1.00554.S & 0.57 $\times$ 0.46 & Noise    \\
36,10,27 ($0^-$) -- 35,9,26 ($0^-$)     & 309.73109 & 340.4 & 5.2  & 2022.1.00554.S & 0.56 $\times$ 0.45 & Blended with CH$_3$COOH \\
22,12,11 ($0^+$) -- 21,11,10 ($0^+$)    & 316.85362 & 144.9 & 5.5  & 2013.1.00061.S & 0.50 $\times$ 0.50 & Noise    \\
\hline
\end{tabular}
\end{center}
\label{tab_archive}
{\bf Notes:} The following criteria were applied: $A_{\rm ij} \geq 5 \times 10^{-4}$~s$^{-1}$,  
$E_{\rm up} \leq 500$~K, a spectral resolution $d\varv \leq 0.5$~km\,s$^{-1}$, an angular resolution 
between 0.3 and 0.7$\arcsec$, and a sensitivity better than 10~mJy\,beam$^{-1}$ for a bin width 
of 1~km\,s$^{-1}$. The angular resolutions are taken from the header of the fits datacubes provided 
on the archive. 
$^a$The transitions at 240.08569~GHz were not listed in the results of ATOMIS but we included them as 
they are in the same spectral window as the line at 240.06964~GHz and have an $E_{\rm up}$ value close 
to the limit of $5 \times 10^{-4}$~s$^{-1}$. 
$^b$ For each pair of close lines, the $g_{\rm up}$ parameters differ significantly despite similar 
$E_{\rm up}$ and $A_{\rm ij}$.
\end{table*}%

After checking the weblog for each project, we downloaded the FITS cubes provided on the ALMA archive for 
all the listed transitions and extracted the spectra at the position studied in PILS 
($\alpha_{\rm J2000} = 16^{\rm h}$32$^{\rm m}$22$\fs$58, $\delta_{\rm J2000} = -24\degr$28$\arcmin$32.8$\arcsec$). 
For the project 2013.1.00061.S, the data that we used are those reduced by Andreu et al. (in prep.) 
with a restoring beam of 0.5$\arcsec$ similar to what has been done for PILS. We could clearly see that 
the continuum was not properly subtracted at the extracted position for two spectral windows of the project 
2022.1.00554.S (296.4 and 309.7~GHz). We corrected it by fitting a one-order polynomial fit on the channels 
that do not seem to show any line emission.

\begin{figure*}
\centering
\includegraphics[width=1.0\linewidth]{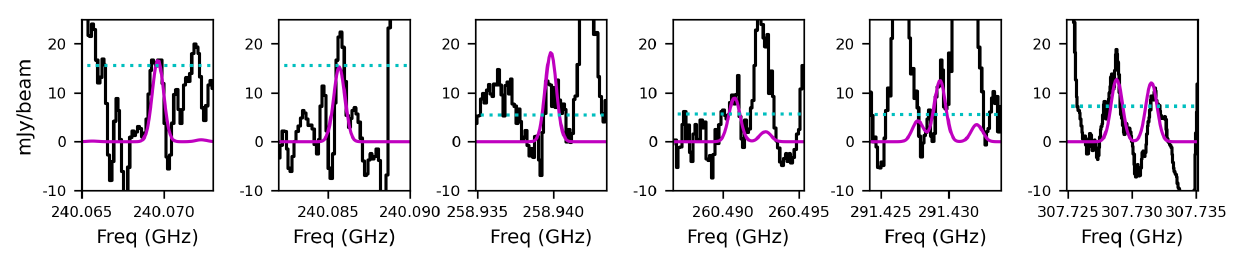}
\caption{\label{fig_det}  Detected lines of 3-hydroxypropenal found in the ALMA archive (in black). 
The model in magenta corresponds to $N = 1.0 \times 10^{15}$~cm$^{-2}$, $T_{\rm ex} = 125$~K, 
a source size of 0.5$\arcsec$ and a FWHM of 1 km\,s$^{-1}$. The cyan dotted line represents the 
3~rms level per 0.33 km\,s$^{-1}$ (1/3 of the line width).}
\end{figure*}

\begin{figure*}
\centering
\includegraphics[width=1.0\linewidth]{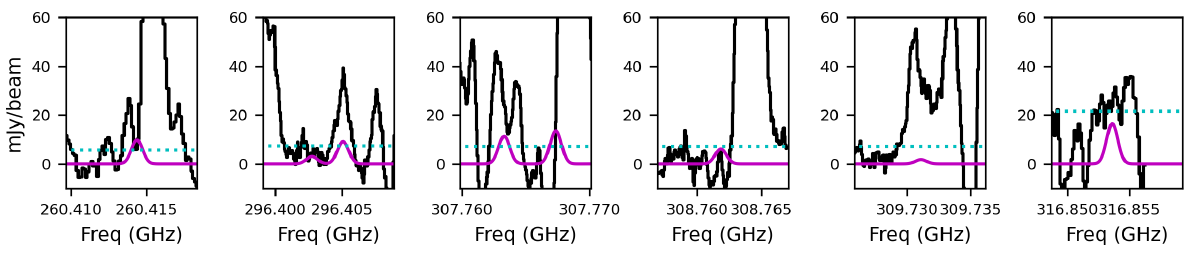}
\caption{\label{fig_undet}  Non-detected lines of 3-hydroxypropenal in the ALMA archive (in black). 
The model in magenta corresponds to $N = 1.0 \times 10^{15}$~cm$^{-2}$, $T_{\rm ex} = 125$~K, 
a source size of 0.5$\arcsec$ and a FWHM of 1 km\,s$^{-1}$. The cyan dotted line represents 
the 3~rms level per 0.33 km\,s$^{-1}$ (1/3 of the line width).}
\end{figure*}


In total, thanks to the ALMA archival data, we detected seven additional lines of 3-hydroxypropenal 
with a signal-to-noise ratio of about 3$\sigma$ or more (see Figure \ref{fig_det}). The model derived with 
the PILS data is in very good agreement with the observations. Figure \ref{fig_undet} shows the lines 
that are not detected. These transitions are not detected either because their predicted intensities 
are too faint compared to the noise level or because they are blended with other species. 
Table~\ref{tab_archive} summarizes for each transition if the line is detected or not with, 
in the latter case, the reason for the non-detection. 
With this new spectroscopic study, the seven additional detected transitions and the good agreement 
of their intensities with the model, the detection of HOCHCHCHO can now be considered as secure. 
The excitation temperature of $\sim$125~K and the column density of $\sim$1.0\,$\times$\,10$^{15}$~cm$^{-2}$ 
of this molecule are also reliable, which provides useful constraints for chemical models 
\citep{3-HP_det_2022}.

\section{Conclusions and outlook}
\label{conclusion}

We have obtained a very large line list of the tunneling-rotation spectrum of 3-hydroxypropenal 
in the course of the present investigation that covers large frequency regions and, more importantly, 
large ranges of quantum numbers. This line list was the basis of an extensive and very accurate 
set of spectroscopic parameters from which the entire tunneling-rotation spectrum of the molecule 
in the ground vibrational state can be derived with great accuracy. 
This new calculation enabled us to settle the issue of the excitation temperature of 3-hydroxypropenal 
in the warm part of the molecular cloud surrounding the protostar IRAS16293~B, as we were able to 
rule out $T_{\rm ex} \approx 300$~K and establish $T_{\rm ex} \approx 125$~K. 
The detection of seven additional lines in the ALMA archive raised the detection from somewhat tentative 
to secure. 
With the new data set, it is now possible to search for 3-hydroxypropenal in star-forming regions 
with considerably higher $T_{\rm ex}$.


\begin{acknowledgements}
This paper makes use of the following ALMA data: ADS/JAO.ALMA\#2013.1.00278.S,  
ADS/JAO.ALMA\#2013.1.00061.S, ADS/JAO.ALMA\#2017.1.01565.S, ADS/JAO.ALMA\#2021.1.00544.S, 
and ADS/JAO.ALMA\#2022.1.00554.S. 
ALMA is a partnership of ESO (representing its member states), NSF (USA) and NINS (Japan), 
together with NRC (Canada) and NSC and ASIAA (Taiwan), in cooperation with the Republic 
of Chile. The Joint ALMA Observatory is operated by ESO, AUI/NRAO and NAOJ.
The work in Lille and Rennes was supported by the Programme National ``Physique et Chimie 
du Milieu Interstellaire'' (PCMI) of CNRS Terre @ Univers with CNRS Physique \& CNRS Chimie, 
co-funded by CEA and CNES.
HSPM acknowledges support by the Deutsche Forschungsgemeinschaft via the collaborative research 
centers SFB~956 (project ID 184018867), subproject B3, and SFB~1601 (project ID 500700252), 
subprojects Inf and A4. 
AC received funding from the European Research Council (ERC) under the European Union's 
Horizon 2020 research and innovation programme (ERC Starting Grant ``Chemtrip'', grant 
agreement no. 949278).
JKJ is supported by the Independent Research Fund Denmark (grant number 0135-00123B). 
Our research benefited from NASA's Astrophysics Data System (ADS).
\end{acknowledgements}


\bibliographystyle{aa} 
\bibliography{3-HP} 

\begin{thebibliography}{33}
\expandafter\ifx\csname natexlab\endcsname\relax\def\natexlab#1{#1}\fi

\bibitem[{{Baba} {et~al.}(1999){Baba}, {Tanaka}, {Morino}, {Yamada}, \&
  {Tanaka}}]{3-HP_tunneling_1999}
{Baba}, T., {Tanaka}, T., {Morino}, I., {Yamada}, K. M.~T., \& {Tanaka}, K.
  1999, \jcp, 110, 4131

\bibitem[{{Baughcum} {et~al.}(1981){Baughcum}, {Duerst}, {Rowe}, {Smith}, \&
  {Wilson}}]{3-HP_rot_etc_1981}
{Baughcum}, S.~L., {Duerst}, R.~W., {Rowe}, W.~F., {Smith}, Z., \& {Wilson},
  E.~B. 1981, J. Am. Chem. Soc., 103, 6296

\bibitem[{{Baughcum} {et~al.}(1984){Baughcum}, {Smith}, {Wilson}, \&
  {Duerst}}]{3-HP_analysis_1_1984}
{Baughcum}, S.~L., {Smith}, Z., {Wilson}, E.~B., \& {Duerst}, R.~W. 1984, J.
  Am. Chem. Soc., 106, 2260

\bibitem[{{Calcutt} {et~al.}(2018){Calcutt}, {J{\o}rgensen}, {M{\"u}ller},
  {Kristensen}, {Coutens}, {Bourke}, {Garrod}, {Persson}, {van der Wiel}, {van
  Dishoeck}, \& {Wampfler}}]{CH3CN_PILS_2018}
{Calcutt}, H., {J{\o}rgensen}, J.~K., {M{\"u}ller}, H.~S.~P., {et~al.} 2018,
  \aap, 616, A90

\bibitem[{{Calcutt} {et~al.}(2019){Calcutt}, {Willis}, {J{\o}rgensen},
  {Bjerkeli}, {Ligterink}, {Coutens}, {M{\"u}ller}, {Garrod}, {Wampfler}, \&
  {Drozdovskaya}}]{CH3CCH_PILS_2019}
{Calcutt}, H., {Willis}, E.~R., {J{\o}rgensen}, J.~K., {et~al.} 2019, \aap,
  631, A137

\bibitem[{{Christen} \& {M{\"u}ller}(2003)}]{aGg-eglyc_rot_2003}
{Christen}, D. \& {M{\"u}ller}, H.~S.~P. 2003, Phys. Chem. Chem. Phys., 5, 3600

\bibitem[{{Coutens} {et~al.}(2023){Coutens}, {Ben Hmida}, \&
  {Glorian}}]{Coutens2023}
{Coutens}, A., {Ben Hmida}, S., \& {Glorian}, J.~M. 2023, in SF2A-2023:
  Proceedings of the Annual meeting of the French Society of Astronomy and
  Astrophysics, 311--312

\bibitem[{{Coutens} {et~al.}(2016){Coutens}, {J{\o}rgensen}, {van der Wiel},
  {M{\"u}ller}, {Lykke}, {Bjerkeli}, {Bourke}, {Calcutt}, {Drozdovskaya},
  {Favre}, {Fayolle}, {Garrod}, {Jacobsen}, {Ligterink}, {{\"O}berg},
  {Persson}, {van Dishoeck}, \& {Wampfler}}]{deuterated_PILS_2016}
{Coutens}, A., {J{\o}rgensen}, J.~K., {van der Wiel}, M.~H.~D., {et~al.} 2016,
  \aap, 590, L6

\bibitem[{{Coutens} {et~al.}(2022){Coutens}, {Loison}, {Boulanger}, {Caux},
  {M{\"u}ller}, {Wakelam}, {Manigand}, \& {J{\o}rgensen}}]{3-HP_det_2022}
{Coutens}, A., {Loison}, J.~C., {Boulanger}, A., {et~al.} 2022, \aap, 660, L6

\bibitem[{{Coutens} {et~al.}(2018){Coutens}, {Willis}, {Garrod}, {M{\"u}ller},
  {Bourke}, {Calcutt}, {Drozdovskaya}, {J{\o}rgensen}, {Ligterink}, {Persson},
  {St{\'e}phan}, {van der Wiel}, {van Dishoeck}, \&
  {Wampfler}}]{H2NCN_PILS_2018}
{Coutens}, A., {Willis}, E.~R., {Garrod}, R.~T., {et~al.} 2018, \aap, 612, A107

\bibitem[{{Endres} {et~al.}(2016){Endres}, {Schlemmer}, {Schilke}, {Stutzki},
  \& {M{\"u}ller}}]{CDMS_2016}
{Endres}, C.~P., {Schlemmer}, S., {Schilke}, P., {Stutzki}, J., \&
  {M{\"u}ller}, H. S.~P. 2016, J. Mol. Spectrosc., 327, 95

\bibitem[{{Firth} {et~al.}(1991){Firth}, {Beyer}, {Dvorak}, {Reeve}, {Grushow},
  \& {Leopold}}]{3-HP_FIR_1991}
{Firth}, D.~W., {Beyer}, K., {Dvorak}, M.~A., {et~al.} 1991, \jcp, 94, 1812

\bibitem[{{Isaacson} \& {Morokuma}(1975)}]{3-HP_ai_1975b}
{Isaacson}, A.~D. \& {Morokuma}, K. 1975, J. Am. Chem. Soc., 97, 4453

\bibitem[{{J{\o}rgensen} {et~al.}(2018){J{\o}rgensen}, {M{\"u}ller}, {Calcutt},
  {Coutens}, {Drozdovskaya}, {{\"O}berg}, {Persson}, {Taquet}, {van Dishoeck},
  \& {Wampfler}}]{PILS_div-isos_2018}
{J{\o}rgensen}, J.~K., {M{\"u}ller}, H.~S.~P., {Calcutt}, H., {et~al.} 2018,
  \aap, 620, A170

\bibitem[{{J{\o}rgensen} {et~al.}(2016){J{\o}rgensen}, {van der Wiel},
  {Coutens}, {Lykke}, {M{\"u}ller}, {van Dishoeck}, {Calcutt}, {Bjerkeli},
  {Bourke}, {Drozdovskaya}, {Favre}, {Fayolle}, {Garrod}, {Jacobsen},
  {{\"O}berg}, {Persson}, \& {Wampfler}}]{PILS_overview_2016}
{J{\o}rgensen}, J.~K., {van der Wiel}, M.~H.~D., {Coutens}, A., {et~al.} 2016,
  \aap, 595, A117

\bibitem[{{Karlstr\"om} {et~al.}(1975){Karlstr\"om}, {Wennerstr\"om},
  {J\'onsson}, {Fors\'en}, {Alml\"of}, \& {Roos}}]{3-HP_ai_1975a}
{Karlstr\"om}, G., {Wennerstr\"om}, H., {J\'onsson}, B., {et~al.} 1975, J. Am.
  Chem. Soc., 97, 4188

\bibitem[{{L{\"u}ttschwager} {et~al.}(2013){L{\"u}ttschwager}, {Wassermann},
  {Coussan}, \& {Suhm}}]{3-HP_exp_tunneling_exc-st_2013}
{L{\"u}ttschwager}, N. O.~B., {Wassermann}, T.~N., {Coussan}, S., \& {Suhm},
  M.~A. 2013, Mol. Phys., 111, 2211

\bibitem[{{Lykke} {et~al.}(2017){Lykke}, {Coutens}, {J{\o}rgensen}, {van der
  Wiel}, {Garrod}, {M{\"u}ller}, {Bjerkeli}, {Bourke}, {Calcutt},
  {Drozdovskaya}, {Favre}, {Fayolle}, {Jacobsen}, {{\"O}berg}, {Persson}, {van
  Dishoeck}, \& {Wampfler}}]{PILS_COMs_2017}
{Lykke}, J.~M., {Coutens}, A., {J{\o}rgensen}, J.~K., {et~al.} 2017, \aap, 597,
  A53

\bibitem[{{Margul{\`e}s} {et~al.}(2023){Margul{\`e}s}, {Coutens}, {Ligterink},
  {Ahmadi}, {Motiyenko}, {Alekseev}, {Vastel}, {Caux}, \&
  {Guillemin}}]{HAN_isos_rot_2023}
{Margul{\`e}s}, L., {Coutens}, A., {Ligterink}, N.~F.~W., {et~al.} 2023,
  \mnras, 524, 1211

\bibitem[{{Margul{\`e}s} {et~al.}(2017){Margul{\`e}s}, {McGuire}, {Senent},
  {Motiyenko}, {Remijan}, \& {Guillemin}}]{HAN_rot_2017}
{Margul{\`e}s}, L., {McGuire}, B.~A., {Senent}, M.~L., {et~al.} 2017, \aap,
  601, A50

\bibitem[{{Margul{\`e}s} {et~al.}(2020){Margul{\`e}s}, {Motiyenko}, \&
  {Demaison}}]{18O-SO2_rot_AuS-red_2020}
{Margul{\`e}s}, L., {Motiyenko}, R.~A., \& {Demaison}, J. 2020, \jqsrt, 253,
  107153

\bibitem[{{M{\"u}ller} {et~al.}(2016){M{\"u}ller}, {Belloche}, {Xu}, {Lees},
  {Garrod}, {Walters}, {van Wijngaarden}, {Lewen}, {Schlemmer}, \&
  {Menten}}]{RSH_ROH_2016}
{M{\"u}ller}, H. S.~P., {Belloche}, A., {Xu}, L.-H., {et~al.} 2016, \aap, 587,
  A92

\bibitem[{{M{\"u}ller} {et~al.}(2023){M{\"u}ller}, {Garrod}, {Belloche},
  {Rivilla}, {Menten}, {Jim{\'e}nez-Serra}, {Mart{\'\i}n-Pintado}, {Lewen}, \&
  {Schlemmer}}]{DMA_rot_2023}
{M{\"u}ller}, H.~S.~P., {Garrod}, R.~T., {Belloche}, A., {et~al.} 2023, \mnras,
  523, 2887

\bibitem[{{M{\"u}ller} {et~al.}(2005){M{\"u}ller}, {Schl{\"o}der}, {Stutzki},
  \& {Winnewisser}}]{CDMS_2005}
{M{\"u}ller}, H. S.~P., {Schl{\"o}der}, F., {Stutzki}, J., \& {Winnewisser}, G.
  2005, J. Mol. Struct., 742, 215

\bibitem[{{Pearson} \& {Drouin}(2005)}]{PrgOH_rot_2005}
{Pearson}, J.~C. \& {Drouin}, B.~J. 2005, J. Mol. Spectrosc., 234, 149

\bibitem[{{Pickett}(1972)}]{RAS_Pickett_1972}
{Pickett}, H.~M. 1972, \jcp, 56, 1715

\bibitem[{{Pickett}(1991)}]{spfit_1991}
{Pickett}, H.~M. 1991, J. Mol. Spectrosc., 148, 371

\bibitem[{{Rowe} {et~al.}(1976){Rowe}, {Duerst}, \&
  {Wilson}}]{3-HP_rot_early-account_1976}
{Rowe}, W. F.~J., {Duerst}, R.~W., \& {Wilson}, E.~B. 1976, J. Am. Chem. Soc.,
  98, 4021

\bibitem[{{Stolze} {et~al.}(1983){Stolze}, {H\"ubner}, \&
  {Sutter}}]{3-HP_rot_Zeeman_1983}
{Stolze}, M., {H\"ubner}, D., \& {Sutter}, D.~H. 1983, J. Mol. Struct., 97, 243

\bibitem[{{Trivella} {et~al.}(2008){Trivella}, {Coussan}, \&
  {Chiavassa}}]{MA-synth_2008}
{Trivella}, A., {Coussan}, S., \& {Chiavassa}, T. 2008, Synth. Commun., 38,
  3285

\bibitem[{{Turner} {et~al.}(1984){Turner}, {Baughcum}, {Coy}, \&
  {Smith}}]{3-HP_analysis_2_1984}
{Turner}, P., {Baughcum}, S.~L., {Coy}, S.~L., \& {Smith}, Z. 1984, J. Am.
  Chem. Soc., 106, 2265

\bibitem[{{Wang} {et~al.}(2008){Wang}, {Braams}, {Bowman}, {Carter}, \&
  {Tew}}]{3-HP_PES_2008}
{Wang}, Y., {Braams}, B.~J., {Bowman}, J.~M., {Carter}, S., \& {Tew}, D.~P.
  2008, \jcp, 128, 224314

\bibitem[{{Zakharenko} {et~al.}(2015){Zakharenko}, {Motiyenko}, {Margul{\`e}s},
  \& {Huet}}]{Lille-spectrometer}
{Zakharenko}, O., {Motiyenko}, R.~A., {Margul{\`e}s}, L., \& {Huet}, T.~R.
  2015, J. Mol. Spectrosc., 317, 41

\end{thebibliography}

\end{document}